%% file: paper.tex
\PassOptionsToPackage{table}{xcolor}

\documentclass[sigconf,9pt,nonacm]{acmart}
\usepackage{blindtext}
\usepackage[english]{babel}

\usepackage[utf8]{inputenc}
\usepackage{tabularx}
\usepackage{booktabs}
\usepackage{graphicx}
\usepackage{multicol}
\usepackage[ruled,linesnumbered,noline,noend]{algorithm2e}
\usepackage{color}
\usepackage{subfig}
\usepackage{balance}
\usepackage{csquotes} 
\usepackage{makecell} 
\usepackage{mathpartir} 

\usepackage{fancyvrb}

\input{vinsyntax}
\input{appearance_hacks}

\usepackage[outline]{contour}
\contourlength{0.2pt}
\contournumber{12}

\usepackage{siunitx}
\usepackage{amsmath}

\usepackage{enumitem}

\newcommand{\dpw}[1]{\textcolor{blue}{\textbf [DPW: #1]}}

\newcommand{\revised}[1]{#1}
\newcommand{\rerevised}[1]{#1}

\newcommand{\lang}[0]{Lucid\xspace}
\newcommand{\OMIT}[1]{}
\newcommand{\EG}{\emph{e.g.}}
\newcommand{\IE}{\emph{i.e.}}

\newcommand{\exprog}[0]{fast rerouter\xspace}

\newcommand{\ns}[1]{\SI{#1}{\nano\second}}
\newcommand{\us}[1]{\SI{#1}{\micro\second}}
\newcommand{\ms}[1]{\SI{#1}{\milli\second}}

\newcommand{\code}[1]{\begin{footnotesize} \texttt {#1} \end{footnotesize}}

\newcommand{\majorheading}[1]{\vspace{8pt}\noindent\textbf{#1}}

\newcommand{\heading}[1]{\vspace{4pt}\noindent\textbf{#1}}

\newcommand{\subheading}[1]{\vspace{2pt}\textit{#1:}}

\newcommand{\topheading}[1]{\noindent\textbf{#1}}

\newcommand{\inl}[1]{\lstinline[style=P4Snip]{#1}}

\newcommand{\relationRule}[4][]{\inferrule*[lab={\sc #2},#1]{#3}{#4}}
\newcommand{\Z}{\ensuremath{\mathbb{Z}}}

\newcommand{\fnwords}[1]{\begin{footnotesize}#1\end{footnotesize}}

\makeatletter
\newcommand{\removelatexerror}{\let\@latex@error\@gobble}
\makeatother

\newcommand\blfootnote[1]{%
  \begingroup
  \renewcommand\thefootnote{}\footnote{#1}%
  \addtocounter{footnote}{-1}%
  \endgroup
}

\acmYear{2021}\copyrightyear{2021}
\setcopyright{acmlicensed}
\acmConference[SIGCOMM '21]{ACM SIGCOMM 2021 Conference}{August 23--28, 2021}{Virtual Event, USA}
\acmBooktitle{ACM SIGCOMM 2021 Conference (SIGCOMM '21), August 23--28, 2021, Virtual Event, USA}
\acmPrice{15.00}
\acmDOI{10.1145/3452296.3472903}
\acmISBN{978-1-4503-8383-7/21/08}

\begin{document}

\title{Lucid: A Language for Control in the Data Plane}

\author{John Sonchack}
\affiliation{
    \institution{Princeton University}
    \city{}
    \country{}
    }
\email{jsonch@princeton.edu}

\author{Devon Loehr}
\affiliation{
    \institution{Princeton University}
    \city{}
    \country{}
    }
\email{dloehr@princeton.edu}

\author{Jennifer Rexford}
\affiliation{
    \institution{Princeton University}
    \city{}
    \country{}
    }
\email{jrex@cs.princeton.edu}

\author{David Walker}
\affiliation{
    \institution{Princeton University}
    \city{}
    \country{}
    }
\email{dpw@cs.princeton.edu}


\input{sections/abstract}
\maketitle

\makeunderscoreletter
\input{sections/intro3}
\input{sections/pisacontrol3}
\input{sections/eventdrivenprogramming}

\input{sections/stateful}
\input{sections/type}

\input{sections/compiling}
\input{sections/eval}

\input{sections/related}

\input{sections/concl}

\bibliographystyle{ACM-Reference-Format}
\bibliography{paper}

\input{sections/appendix}

\end{document}

%% file: vinsyntax.tex
\usepackage{syntax}     

\newcommand{\makeunderscoreletter}{\catcode`\_=11}
\newcommand{\makeunderscoreactive}{\catcode`\_=\active}

\makeatletter
\def\reuseopen{%
\relax%
\syn@assist%
  {\color{gray}$\langle$\normalfont\itshape}%
  {\act_{\@uscore.}}%
  >%
  \syntright%
  {\relax}%
}
\makeatother

\usepackage{listings}
\usepackage{solarized}      
\usepackage[normalem]{ulem} 
\usepackage{etoolbox}       

\newbool{rm}
\newbool{add}

\newcommand{\dyncolor}[2]{%
\ifbool{add}%
{\textbf{\textcolor{#1}{#2}}}%
{\ifbool{rm}{%
\sout{\textcolor{#1}{#2}}}%
{\textcolor{#1}{#2}}}%
}

\lstdefinelanguage{Dptlang}
{
  morekeywords={packet, entry, exit, support, event, handle, int, if, generate, hash, const, memop, fun, return},
  morecomment=[l]{//},
  morestring=[b]",
  moredelim=*[is][\dyncolor{scyan}]{\%}{\%},
  moredelim=*[is][\dyncolor{sblue}]{\!}{\!},
  moredelim=*[is][\dyncolor{purple}]{|}{|},
  moredelim=**[is][\color{sbase3light}]{<>}{<>}
}
\lstdefinestyle{DptBlock}
{
  moredelim=*[is][\dyncolor{sblue}]{\^}{\^},
  morecomment=[l]{//},
  morestring=[b]",
  keywords=[1]{global, memop, fun, handle, event},
  keywordstyle=[1]\dyncolor{blue},
  keywords=[2]{int},
  keywordstyle=[2]\dyncolor{syellow},
  keywords=[3]{return, generate, const, new, packet, entry, exit, support, if, else},
  keywordstyle=[3]\dyncolor{smagenta},
  keywords=[4]{Array},
  keywordstyle=[4]\dyncolor{scyan}
}
\lstdefinestyle{SmallDptBlock}{
  basicstyle=\scriptsize\ttfamily\color{sbase02},
  style=DptBlock
}

\lstdefinestyle{DptSnip}
{
  language=Dptlang,
  backgroundcolor=\color{white}, 
  frame=none  
}

\lstdefinelanguage{P4lang}
{
  keywords=[1]{Register, RegisterAction, RegisterActions, Registers, action, table},
  keywordstyle=[1]\dyncolor{blue},
  keywords=[2]{struct, inout, apply, header_t, metadata_t, import, control, in, out, bit, if, else, const},
  keywordstyle=[2]\dyncolor{syellow},
  morecomment=[l]{//},
  morestring=[b]",
  moredelim=*[is][\dyncolor{scyan}]{\%}{\%},
  moredelim=*[is][\dyncolor{sblue}]{\^}{\^},
  moredelim=**[is][\color{sbase3light}]{<>}{<>}
}

\lstdefinestyle{P4Block}
{
  language=P4lang
}

\lstdefinestyle{SmallP4Block}
{    
  basicstyle=\scriptsize\ttfamily\color{sbase02},
  language=P4lang
}
\lstdefinestyle{P4Snip}
{
  language=P4lang,
  backgroundcolor=\color{white}, 
  frame=none  
}

\lstdefinestyle{SmallP4Snip}
{
  basicstyle=\scriptsize\ttfamily\color{sbase02},
  language=P4lang,
  backgroundcolor=\color{white}, 
  frame=none  
}

\lstdefinestyle{PsuedoSnip}
{
  language=C++,
  moredelim=*[is][\dyncolor{sblue}]{\!}{\!},
  morekeywords={bit},
  backgroundcolor=\color{white},
  frame=none
}

%% file: appearance_hacks.tex
\definecolor{pennred}{RGB}{149,0,26}
\usepackage{tikz}
\newcommand{\circled}[2][]{%
  \tikz[baseline=(char.base)]{%
    \node[shape = circle, fill = pennred, draw, inner sep = 0pt]
    (char) {\phantom{\ifblank{#1}{#2}{#1}}};%
    \color{white}\footnotesize
    \node at (char.center) {\makebox[0pt][c]{#2}};}}
\robustify{\circled}

\newcolumntype{Y}{>{\centering\arraybackslash}X}

%% file: sections/abstract.tex
\begin{abstract}

Programmable switch hardware makes it possible to move fine-grained control logic inside the network data plane, improving performance for a wide range of applications. However, applications with integrated control are inherently hard to write in existing data-plane programming languages such as P4. This paper presents Lucid, a language that raises the level of abstraction for putting control functionality in the data plane. Lucid introduces abstractions that make it easy to write sophisticated data-plane applications with interleaved packet-handling and control logic, specialized type and syntax systems that prevent programmer bugs related to data-plane state, and an open-sourced compiler that translates Lucid programs into P4 optimized for the Intel Tofino. These features make Lucid general and easy to use, as we demonstrate by writing a suite of ten different data-plane applications in Lucid. Working prototypes take well under an hour to write, even for a programmer without prior Tofino experience, have around 10x fewer lines of code compared to P4, and compile efficiently to real hardware. 
\rerevised{
In a stateful firewall written in Lucid, we find that moving control from a switch's CPU to its data-plane processor using Lucid reduces the latency of performance-sensitive operations by over 300X.}

\end{abstract}

%% file: sections/intro3.tex
\section{Introduction}
In the early days of Software-Defined Networking (SDN), controller applications changed network behavior by updating the match-action rules that switches use to forward packets.  Unfortunately, many interesting applications needed the switches to direct packets to the controller (\EG, to learn about new flows and install new rules in response), causing latency, overhead, and security vulnerabilities.  In addition, writing these applications was tricky, since programmers had to reason about possible inconsistencies between the switches and the controller due to delays in installing new rules. As a result, few of these dynamic controller applications saw any significant deployment in practice. \blfootnote{Lucid is available at: \url{https://github.com/princetonUniversity/lucid}}



The emergence of programmable data-plane hardware has the potential to change all that, by making it possible to move fine-grained control logic into the forwarding engines of individual switches. Modern hardware, \IE, a PISA pipeline~\cite{bosshart2013forwarding}, supports not only flexible parsing and manipulation of packets, but also updates to persistent state (such as register memory) that can be used to affect the handling of future traffic.


Consider, for example, a stateful firewall that protects an enterprise from unsolicited traffic. By default, packets sent by external hosts are dropped. Upon receiving outbound packets, the switch state is updated to permit return traffic from the destination; after a period of inactivity, the switch returns to dropping such packets.  Having a controller handle these ``events''---the arrival of the first packet and the timeout after inactivity---introduces latency and overhead, and the subtle risk that return traffic starts arriving before the data plane is updated to permit it. Implementing this logic directly in the data plane reduces reaction time, and
avoids the need for synchronization between controller and data plane.

The stateful firewall is just one of many examples. Figure~\ref{fig:usecases} presents several other applications, along with their state and their data-plane and control-plane components. Past researchers have demonstrated that many of these applications, including load balancers~\cite{alizadeh2014conga,katta2016hula,hsu2020contra}, routers~\cite{hsu2020contra}, and telemetry systems~\cite{starflow} benefit substantially from data-plane implementations.

\begin{figure*}[!t]
\centering
\footnotesize
\setlength{\tabcolsep}{2pt}
\begin{tabularx}{\linewidth}{ X X X X } 
\toprule
 \textbf{Application} & \textbf{Data Plane Operations} & \textbf{Control Plane Operations} & \textbf{Shared State} \\
 \midrule
 \textbf{Router} 
    & get next hop 
    & update route 
    & forwarding database \\  
 \textbf{Fault-Tolerant Router} 
    & get online path 
    & probe link, reroute 
    & link and route statistics \\
\textbf{MAC Learner} 
    & get output port 
    & learn, age, STP 
    & MAC table \\
\hline
 \textbf{Flow\, Load\, Balancer} 
    & count packets, get\,path 
    & set path 
    & flow\,stats,\,paths \\
 \textbf{Flowlet\, Load\, Balancer} 
    & set, get path 
    & age flowlets 
    & flowlet hash table \\
  \textbf{NAT}  
    & translate address 
    & allocate, pin, free 
    & address map and pool \\
  \textbf{TCP\,Migration} 
    & adjust ack number 
    & migrate flow 
    & ack offset table \\
\hline  
 \textbf{Stateful\, Firewall} 
    & check flow 
    & add flow 
    & allowed\,flow\,set \\ 
 \textbf{Probabilistic Firewall} 
    & check, add flow 
    & rotate (age) 
    & Bloom filters \\
 \textbf{Distributed Probabilistic FW} 
    & check flow 
    & add,\,age,\,sync 
    & replicated Bloom filters \\ 
\hline
 \textbf{Sketch-based Query} 
    & update sketch 
    & reset, decode 
    & approximate data structure \\ 
 \textbf{Telemetry Cache} 
    & append packet record 
    & evict, free 
    & per-flow log\\ 
\bottomrule
\end{tabularx}
\vspace{-4mm}
\caption{Example applications in groups: (1) Routing, (2) Connection-oriented services, (3) Security, and (4) Telemetry.}
\vspace{-4mm}

\label{fig:usecases}
\end{figure*}

Despite this, writing applications that do network control inside of PISA data planes is incredibly difficult. Languages like P4 offer the abstraction of single-threaded packet processing that does little (if any) state management. However, applications like the stateful firewall require multiple threads, one for packet handling (\EG, forwarding permitted packets and dropping the rest) and several more for control operations that manage local state (\EG, updating the rules in response to both flow arrivals and timeouts). 

A programmer's first challenge is simply expressing control with packet processing abstractions, which requires using disjoint and low-level primitives such as parsers, match-action units, and packet recirculation. For more sophisticated control, such as distributed route computation or a delayed scan for inactive flows, programmers must also carefully orchestrate PISA components that are outside the scope of data-plane languages, such as programmable queues and packet generators. 

Beyond this, there is a second equally daunting challenge: operating on persistent state (registers) in the data plane. The demands of line-rate constrain the Arithmetic Logic Units (ALUs) that operate on registers in all sorts of nuanced ways. Control operations can be complex and run up against these limits. Further, a PISA processor is a pipeline where each register is associated with a single stage. Multiple threads of control that share state must always access the underlying registers in a consistent order. 

None of the above constraints are enforced in the stateful primitives of data-plane languages. So when compilation fails because of inevitable programmer error, the failure often occurs in a target-specific backend that is ill-equiped to provide meaningful source-level programmer feedback. Faced with this, the programmer must resort to a painful trial-and-error process of rewriting the program and grappling with cryptic compiler errors, never sure when the compiler will finally yield to the next tweak of the program.

\majorheading{\lang:  Simple, Event-driven Data-Plane Programming.}
This paper introduces \lang, a high-level programming language for implementing control applications in PISA data planes. 

\lang programs are organized as multiple collaborating components located either on a single switch or distributed across many switches in a network. 
In \lang, programmers program with high-level, abstract \emph{events} and \emph{handlers}. Each event is named and carries user-specified data, while its associated handler defines the atomic stateful computation to perform when an event occurs. An event could be a packet to process, a request to install a firewall entry, or a probe from a neighboring switch to report a link's status. 

Events are also associated with \emph{times} and \emph{locations} to facilitate coordination between control operations among different switches, possibly with a delay.
Programmers can write sophisticated control logic  without having to worry about the low-level details of custom packet formats, parsers and deparsers, or having to ``roll their own'' mechanisms for buffering and delayed information processing.

\lang's event-based abstractions for structuring applications and coordinating control are complemented by a careful ``correct-by-construction'' approach to stateful operations. Rather than allow programmers to write event handler code that operates on state in arbitrary ways, but may produce arbitrary failures in a PISA compiler's backend,
\lang introduces a persistent array abstraction that is carefully designed to rule out illegal constructions.
This abstraction is supported by domain-specific syntactic constraints and a novel type
system:

\begin{itemize}[leftmargin=*]
    \item \textbf{Syntactic constraints}: We design a sublanguage of \emph{memops}, stateful operations that can execute in a single ALU of a PISA switch. Memop definitions that cannot fit in a single ALU are rejected, and source-level error messages point out exactly where any such mistakes occur, making it easier for programmers to understand how and why they must change the processing of an individual control operation. 

    \item \textbf{Types and effects}: We develop a novel 
    \emph{ordered type-and-effect system} that limits the way programs interact with persistent memory. Our type system tracks the order in which handlers access registers and provides actionable source-level feedback when there are inconsistencies. This feedback helps programmers quickly identify control operations that must be changed (\EG, decomposed into multiple simpler operations).
\end{itemize}
\noindent

Together, \lang's event-based abstractions and carefully-designed stateful interface give programmers a natural and modular way to express data-plane applications that interleave packet processing with ongoing control operations \emph{and} help them navigate the difficulty of programming complex switch hardware. 

The \lang compiler shows how to map the above ideas to real hardware---the Intel Tofino. Our compiler analyzes \lang programs and translates valid programs into Tofino-compatible \texttt{P4\_16}.  It also optimizes them to reduce pipeline resource requirements.




We evaluate \lang by implementing a diverse set of applications including a stateful firewall, 
fault tolerant router, self-driving DNS protection service, telemetry cache, and more. All programs had 5-10X fewer lines of code than their P4 equivalents and, due to the \lang compiler's optimizations, utilize the Tofino's limited pipeline stages efficiently. In writing these applications, we find that \lang enables high developer productivity and presents a low barrier of entry to high-speed data-plane programming. Several applications were written by a PhD student who had never worked with the Tofino before (one of the authors, who worked only on the language front end). They used \lang to implement a variety of interesting prototype applications, all in well under an hour. \rerevised{In our experience, it often takes students days or even weeks of debugging to get equivalent P4 programs to compile and use resources efficiently.}


Finally, a case study of a stateful firewall written in \lang demonstrates the performance benefit of integrating latency-sensitive network control in the data plane. \rerevised{We find that flow installation time is 300X lower for data-plane integrated control, compared to remote control from the switch CPU.}


\majorheading{Summary.} The main contribution of this paper is the design of \lang, a high-level language for stateful, distributed data-plane programming. More specifically, this paper: 

\begin{itemize}[leftmargin=*]
    \item motivates data-plane integrated control and identifies the enabling mechanisms in PISA processors (Section~\ref{sec:integratedcontrol});
    \item introduces event-based abstractions that naturally generalize both packet and control processing (Section~\ref{sec:events}); 
    \item designs an interface to data-plane state that uses syntactic constraints and a novel type system to identify programs ill-suited for the underlying hardware (Sections~\ref{sec:stateful} and~\ref{sec:type});
    \item describes an open-source, optimizing compiler targeting the Intel Tofino (Section~\ref{sec:compiler}); and 
    \item evaluates \lang, demonstrating that it is general, easy to use, and compiles efficiently to real hardware (Section~\ref{sec:eval}).
\end{itemize}

\revised{
\noindent\textbf{Ethics.} All user data in this work are from paper authors.
}

%% file: sections/pisacontrol3.tex
\section{Integrated Data-Plane Control}
\label{sec:integratedcontrol}
\begin{figure}[t]
\includegraphics[width=0.95\linewidth]{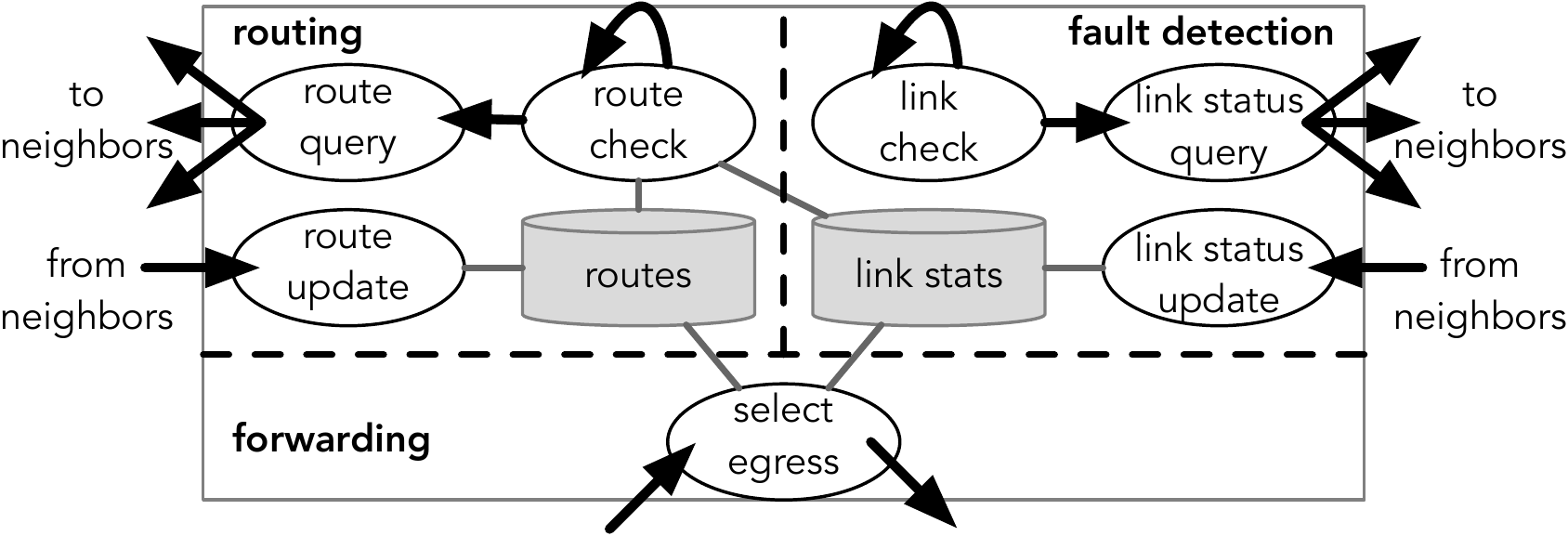}
\vspace{-4mm}
\caption {
\exprog architecture. Circles represent operations, with arrows for (possibly delayed) control flow. Shaded objects are data structures.}
\vspace{-4mm}
\label{fig:router} 
\end{figure}




Many network services benefit from \emph{integrated data-plane control}, \IE, 
the placement of control operations in the data plane's packet
processing hardware rather than in servers or switch management CPUs.
In this section, we illustrate the benefits of integrated data-plane control, and the enabling hardware
mechanisms, with a driving example: the \emph{\exprog}, a fault-tolerant forwarder that detects and dynamically routes around failed links. 

The \exprog (see Figure~\ref{fig:router}) consists of three key components. 

\topheading{Forwarding.} The \exprog looks up a next hop for each packet in an
associative array based on its destination. Before forwarding, 
the program checks a second data structure to determine if the next hop is
still reachable. If not, it may trigger rerouting. 

\topheading{Fault detection.} Concurrent with forwarding, a \exprog node also 
regularly pings all of its directly connected neighbors to determine if they
are still reachable.

\topheading{Rerouting.} Interleaved with the above, a reroute operation in the 
\exprog queries all of its neighbors to find the next hop with the lowest 
route length. This can be triggered by a packet with no next hop 
or a periodic route table scan.




\revised{

\subsection{Motivation: Low Latency Control}

Low latency is a primary motivator for data-plane integrated 
control in many applications. 

\begin{itemize}[leftmargin=*]

\item{\textbf{Fault tolerance services}, such as the
\exprog and F10~\cite{f10}, 
detect and mitigate failures in the data plane to 
minimize reaction time and therefore disruption to traffic.}

\item{In \textbf{5G mobile cores}~\cite{turboepc}, integrating
signal handling operations into the data plane reduces latency by up to 98\%, 
enabling faster connection setup and migration for users.}

\item{\textbf{Load balancers}~\cite{katta2016hula, alizadeh2014conga} 
with control loops in the data plane can react faster to congestion events. This, in turn, improves end-to-end application performance.}

\item{\textbf{Security services} that operate on flows, such as DDoS defense 
systems~\cite{jaqen} and stateful firewalls (Section~\ref{ssec:sfw}), 
integrate control into the data plane to block threats or authorize 
trusted flows with less latency.}

\end{itemize}

\topheading{Root causes.}
In most cases, data-plane integration reduces 
latency by \emph{eliminating communication overheads}.
Consider the \exprog as an example. Detecting and routing around a link
failure requires at least two rounds of messages between a switch and its
neighbors: One round to determine that a next hop has failed and a second 
round to identify an alternate next hop. If the \exprog's control operations ran
as a Linux application on the switch's CPU, the operating system itself would
add around \us{400} of latency because unidirectional messaging between Linux
socket endpoints takes around \us{100} end-to-end~\cite{stackmap}.
However, a version of the \exprog with control in the data plane would completely avoid
this overhead, along with others due to control-related
middleware~\cite{p4runtime, onos}. For example, sending a message (i.e., a packet)
from a switch's data-plane processor to its neighbor takes around \us{1}, and is bound only
by the propagation and queueing delays of the physical hardware. 
Data-plane integration also eliminates the communication overheads between
a switch's data-plane and management processors (\EG, PCIe
latency~\cite{pcieperf}). These overheads dominate in single-node applications
with simpler control operations, such as in a stateful firewall
(Section~\ref{ssec:sfw}).

\subsection{PISA Programmable Packet Processing}

This paper focuses on 
PISA (Protocol Independent Switch Architecture) processors. PISA is a compelling
data-plane architecture for three reasons: First, it is programmable; second, it
is a generalization of real-world chips, primarily the Intel Tofino,
and third; it processes packets at a high and guaranteed line rate (one packet per clock). Given a platform-specific minimum packet size, a PISA processor can sustain a workload
that saturates \emph{all} ports simultaneously.



} 

\begin{figure}[t]
\includegraphics[width=0.90\linewidth]{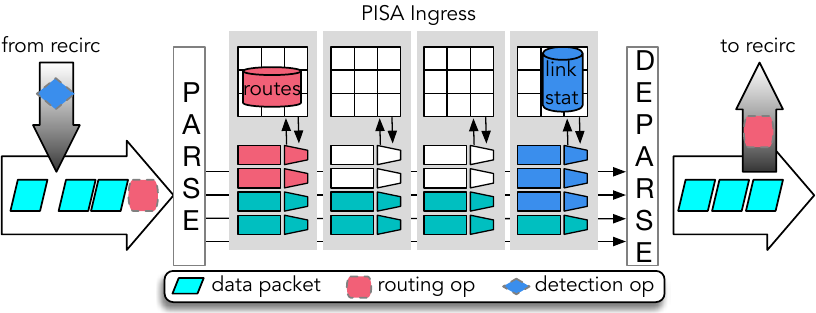}
\vspace{-4mm}
\caption {Interleaving the \exprog's control operations and packet processing in a PISA switch.}
\vspace{-4mm}
\label{fig:execmodel} 
\end{figure}

The core of a PISA processor, illustrated in Figure~\ref{fig:execmodel}, is a
programmable line-rate match-action pipeline. Line rate demands a
tightly synchronized, \emph{feed-forward design}: Each pipeline stage has a
throughput of one packet per clock, and packets only ever move forward through
the pipeline. Instruction-level parallelism is also critical for line
rate. A packet's header moves through each stage in parallel, as a vector.
When the packet header enters a stage, ternary (TCAM) and exact
(hash + SRAM) match-action tables evaluate it to feed ALU vectors with
instructions to modify header fields. The ``programmable'' aspect of the
pipeline is the capability to set table layouts and instructions at compile
time, and set table entries from a management CPU at run time.




Stages also have stateful ALUs (sALUs) for updating local SRAM \emph{register arrays}. Each stateful ALU can read from a single address in
SRAM, perform limited computation, and write back to SRAM or modify metadata associated with the packet. 

\revised{
The ingress pipeline can direct packets to  an egress port or, optionally, a recirculation port that brings the packet back to the start of the pipeline for additional processing. A recirculation port typically has the same bandwidth as a single front-panel port and shares the pipeline's packet-processing bandwidth, so only a fraction of packets can be recirculated without limiting throughput.

Finally, real-world PISA processors also include platform-specific and semi-programmable ``support engines'' outside of the core pipeline. The Intel Tofino, which this paper focuses on, includes four such engines: (1) a multicast engine to copy packets, (2) a queue manager to shape flows, (3) a packet generator for spawning packets, and (4) configurable MAC blocks that can dynamically pause queues based on Priority Flow Control (PFC) frames.
} 

\subsection{Packet-driven Control Operations}


We know that packet processing maps well to a 
PISA chip~\cite{bosshart2013forwarding}, but how do we use it for more general control tasks?
The key is to break control tasks down into atomic operations driven by the
arrival of packets---when an ordinary data packet arrives, its presence and
path through the switch can trigger execution of control operations that occur
alongside regular forwarding operations.  This can work well if control operations align with data packet arrival and are simple enough to fit in a single pass through a PISA switch.  

But what if control operations are complex and their
execution depends on one another rather than on the arrival of ordinary data
packets? For example, Figure~\ref{fig:router} sketches the control structure of
the fast rerouter.  Complex control operations like routing or fault detection
can be decomposed into simpler units---route updates, route checks and route
queries in the case of the routing component, for instance.  And each of those
operations can be performed in a single pass through a PISA pipeline.
To ensure these operations occur at the appropriate cadence, it is 
possible to design, generate and parse new \emph{synthetic} control packets 
to initiate execution of these control operations at the appropriate time.

\subsection{Persistent State}
Typically, the goal of control operations is to update state that affects the
processing of subsequent packets. For example, reroute operations set entries in an array that determines where future packets are forwarded. A
PISA pipeline stores this state in its stage-local SRAM banks. Control (and
packet) operations read and write the state atomically using stateful ALUs. Atomicity means that a stateful operation can only update a single word of memory and perform computation that is simple enough to execute in a single
instruction. 

For more complex stateful operations, we can use multiple SRAM banks or stages. For example, when the \exprog forwards a packet, it looks
up a next hop from an array in one stage, then uses an array in a subsequent stage to determine if the next hop is still active. Conveniently, multi-stage stateful operations are still atomic and line rate because of the
feed-forward architecture of a PISA pipeline~\cite{domino}. But this comes with a programmer challenge because it forces all packet and control operations to access state in the same order.




\subsection{Control Threads via Recirculation} 
Some control operations far exceed the resource limits of a PISA pipeline.
For instance, in the \exprog, checking the status of one link is a simple
computation.  But how do we check the status of an entire table of links?
In general, \emph{maintenance tasks} that require iteration over large tables
or sets of values to identify stale or erroneous entries may appear impossible to support. 
However, we can use recirculation for \emph{serial} processing, by recirculating a control packet multiple times to perform one part of its task in each pass, or we can use it for \emph{parallel} processing, by recirculating multiple control packets back-to-back, each operating on a different entry. 

A potential concern is that recirculation for control consumes bandwidth that
could be used for packet processing. This is possible, but for many
applications, overhead is low because control operations, even low-latency and
fine-grained ones, are infrequent compared to data-plane packets. 
\rerevised{For example,
consider the \exprog on a 1 Ghz PISA with 128 ports. It detects failed links by
serially scanning the link status table with a control packet that
recirculates once per \us{}. The recirculation throughput, 1 million
packets/s, is only 0.1\% of the pipeline's bandwidth. Even though overhead is low, it still checks each port often, once per \us{128}.}


\subsection{Scheduled Control via Support Engines}
Of course, we may not always \emph{want} control threads to operate at the highest possible rate, or, for that matter, at the same switch. 
%
This brings us to the last piece of the puzzle: how can a PISA processor schedule the \emph{place} and \emph{time} where control operations execute? 

\heading{Place.} Changing \emph{where} an operation executes is straightforward, assuming that switches have addresses. Since control operations are processed like packets, a switch can schedule an operation at another location by encapsulating the corresponding control packet in an appropriately addressed frame and 
forwarding it just like any other packet. With line-rate multicast engines, we can even schedule an operation at \emph{multiple} locations (such as the \exprog pinging all neighboring switches) in a single step.

\revised{
\heading{Time.} Changing \emph{when} an operation executes is harder. Essentially, we need to buffer a control packet for some amount of time. A design for a generic PISA processor could buffer it in a register array along with the time at which it should be executed, and then scan the array periodically to find operations ready to execute. However, this approach could consume a large number of stateful ALUs. An alternative is to simply recirculate the control packet repeatedly until it is ready to execute, but this consumes recirculation bandwidth. A more efficient design, specialized for the Intel Tofino, is to use a dedicated queue for delayed control operations, which is paused and periodically released using PFC pause frames from the Tofino's packet generator. 

}

\subsection{Masters of Complexity}

To many a hacker, our discussion of control in the data plane may sound glorious.
All one needs to do is:

\begin{itemize}[leftmargin=*]
    \item break complex control operations into simpler ones;
    \item create formats, parsers, and deparsers for new synthetic packets to
    drive control operations where necessary; 

    \item mind the constraints on stateful ALUs and per-stage computation; 
    \item ensure that control operations access state in a consistent order; 
    \item manually recirculate packets for multi-step operations, 
    carefully interleaving the processing of recirculated and 
    data packets; 
    \item learn how to program the support engines outside the 
    language of your programmable switch; and 
    \item implement primitives for delay and distribution of control.

\end{itemize}

What's not to like?  Everyone loves rolling multiple low-level resource constraints and a couple of different programming paradigms around in their head, before they even get to considering the high-level logic of their application!  
And yet, in our experience, expert programmers can spend weeks and hundreds or
thousands of lines of code developing network control applications with relatively simple high-level ambitions.

Hence, rather than embracing this challenge to master complexity, we propose a
better way: higher-level abstractions, static analysis, and automatic generation of low-level code. As the coming sections will show, these mechanisms together reduce programmer time from days to hours and lines of code by a factor of 10.

%% file: sections/eventdrivenprogramming.tex
\section{Event-Driven Programming}
\label{sec:events}



To create control applications for PISA switches, programmers currently must implement many low-level mechanisms by hand. In many ways, it is reminiscent of writing a distributed system without basic operating system services. 
Consider the challenge of adding the \exprog's \emph{route query} control operation to a basic forwarding program written in P4. We must define a route query header along with parsers and deparsers; adjust the control flow to branch on that header's presence (in addition to existing branches); serialize generated queries into event packets; and finally configure the multicast engine to broadcast these packets to all neighbors.
%
All of this effort only gets us to the point where we can \emph{begin} to implement the interesting logic, e.g., the P4 that generates and responds to route queries.

\subsection{Event-based \lang Abstractions}


%
The main idea behind \lang's core abstractions is
to unify packet processing with control operations through
intuitive primitives for coordinating \emph{when} and \emph{where} events execute.

\heading{Events.}
\lang abstracts both control operations and data packets as \emph{events}. Every event consists of a four-tuple containing (1) a name, (2) carried data, (3) a time, and (4) a place. 
Events give programmers a way to structure multi-threaded programs that is missing from P4 and other existing data-plane languages. For example, the routing component of the \exprog has three events, corresponding to its three operations in Figure~\ref{fig:router}. 
\begin{lstlisting}[style=DptBlock]
event route_query(int sender_id, int dst);
event route_reply(int sender_id, int dst, int pathlen);
event check_route(int dst);
\end{lstlisting}

Events are also a high-level abstraction for application-layer messages. The \inl{route_query} event is a request that switch \inl{sender_id} sends to its neighbor, asking for the length of its path to \inl{dst}. A \inl{route_reply} is a response to a query. Finally, \inl{check_route} is an instruction \emph{that a switch sends to itself} to check whether the route to \inl{dst} has failed. 


\heading{Handlers.} 
In a \lang program, all computation happens in a handler. 
A handler specifies what happens, such as a control operation or a packet-processing function, when a switch gets an event. 
Each handler compiles to a slice of parallel tables, ALUs, and stateful ALUs, and executes in a single pass through the match-action pipeline. Although the low-level implementation is complex, the high-level language for writing handlers is simple and expressive. For example, here is the \inl{route_query} handler from the \exprog.





\begin{lstlisting}[style=DptBlock]
handle route_query(int sender_id, int dst) {
    int pathlen = get_pathlen(dst);
    event reply = route_reply(SELF, dst, pathlen);
    generate Event.locate(reply, sender_id);
}
\end{lstlisting}

The handler runs on a switch when its neighbor \inl{sender_id} schedules a \inl{route_query} event to execute on it. The handler looks up the length of the path to \inl{dst} from a persistent array (\inl{pathlens}), and communicates the result to \inl{sender_id} by scheduling a \inl{route_reply} event to execute there. 
The programmer can implement all the logic for a route query (including the \inl{get_pathlen} function)
while only writing \emph{roughly the number of lines it would take merely to declare a route query header in P4.}

\heading{Event generation.}
As the \inl{route_query} example also shows, handlers not only perform some computation in response to an event/packet, but can also generate events to trigger additional future computation. 
This abstraction of time is powerful because it lets us break complex control operations up into a thread of events that executes over a period of time. 

For example, in the \exprog, we implement the thread that periodically scans the status of every route request as a recursive event handler. The handler checks the status of a route at a certain position in the routing table, and then (recursively) generates another event to check the next position. As another example, the stateful firewall application (Section~\ref{ssec:sfw}) uses an event that recurses a bounded number of times to implement the insert operation of a cuckoo hash table in the data plane.

\heading{Event combinators.}
Event combinators let a handler change \emph{when} and \emph{where} an event that it generates will execute. 

The \textbf{locate} combinator, which the \inl{route_query} example uses, lets a handler specify where an event will be generated. Similar to a \code{send} system call in Linux, the locate combinator provides a simple abstraction for unicast communication. \lang also provides a multicast locate combinator for group communication.


The \textbf{delay} combinator changes \emph{when} an event is executed. This combinator makes it easy to pause persistent computations, and hence resembles the Linux \code{sleep} system call. The \exprog uses the delay combinator to control the rate at which it pings its neighbors and also the rate at which it scans its routing table.

\subsection{Data-Plane Event Scheduler}
\label{ssec:sched}

\begin{figure}[t]
\begin{lstlisting}[style=DptBlock]
event a(); event b(); event c(); const group GRP = {2, 3};
handle a() {
	generate b();
	mgenerate Event.delay (Event.locate (c(), GRP), 10ms);
}
\end{lstlisting}
\vspace{-1mm}
\includegraphics[width=0.95\linewidth]{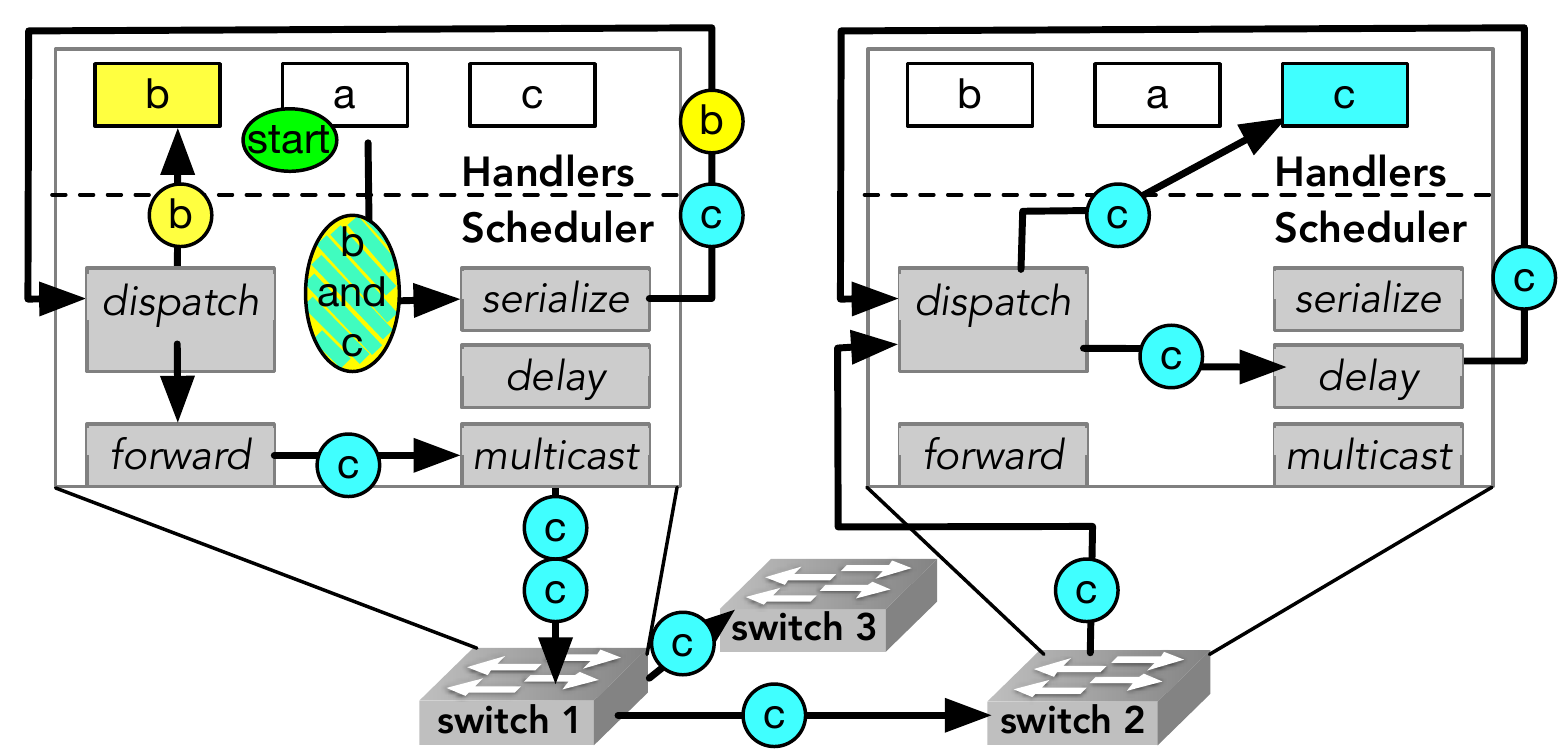}
\vspace{-4mm}
\caption{Event scheduling for a simple program.}
\vspace{-4mm}
\label{fig:evprog}
\end{figure}

\lang realizes the abstractions for event-based distribution and communication using an \emph{event scheduling library} that is inlined into a \lang program. As Figure~\ref{fig:evprog} shows, the library sits logically between a \lang application and the underlying network, filling the role of a lightweight operating system. \revised{The library is mostly comprised of hand-written code that is Tofino specific.}



We describe the main components by following Figure~\ref{fig:evprog}, beginning with the execution of handler \inl{a} at switch 1. This generates two events, \inl{b} and \inl{c}, by removing the \inl{a} event header from the packet and attaching event headers for both \inl{b} and \inl{c}.

\heading{Event serialization.}
After the ingress pipeline finishes, \lang's serializer transforms the single packet with headers for \inl{b} and \inl{c} into \emph{serialized event packets}, one for each event. 

First, the serializer uses the switch's multicast engine to create one copy of the packet for each event. When a copy arrives at the egress pipeline, it has headers for both \inl{b} and \inl{c}. The event serializer deletes one header from each copy, using a clone ID field that the Tofino provides as metadata.  



\heading{Event dispatching.}
The event serializer sends the event packets for \inl{b} and \inl{c} to the switch's recirculation port. When these packets re-enter the ingress pipeline, the event data is extracted by a \lang-generated parser and passed to an \emph{event dispatcher}, an ingress match-action table that performs one of three actions based on an event's \emph{location} and \emph{delay}.

\subheading{Non-local events} If the event's location is not the current switch, the dispatcher calls a user-configured forwarding table to select an output port or multicast group for the event. In the example program, the dispatcher at switch 1 sets a multicast group for event packet \inl{c}, sending it to switches 2 and 3. 

\subheading{Delayed local events} For events that are destined to the local switch, but with a delay $> 0$, the dispatcher calls a \emph{delay} function. In the example, switch 2 initially delays event \inl{c}. 

\subheading{Processable events} When an event's delay is $0$ and its location is the current switch, the dispatcher applies the compiler-generated tables that implement the event handlers. In the example program, the dispatcher at switch 1 will do this for \inl{b} as soon as it arrives.

\heading{Implementing delay.}
Delay is the most sophisticated function in the scheduler. \lang implements this with  \emph{pausable egress queues}. Events to delay are placed in a special ``delay queue'' of the recirculation port. The queue is paused most of the time and unpaused at a regular interval to release packets, e.g., once every \us{100}. When events exit the queue, a table in egress updates their delay parameter based on their queue time. The packets recirculate and repeat until their delay is 0.
PFC (Priority Flow Control) packets let the event scheduler time the queue. We send a stream of packets into the pipeline that consists of pairs of PFC packets at a low, constant rate. The first PFC packet in a pair unpauses the queue to let event packets out, while the second one repauses it. The PFC stream can be generated by either the pipeline's packet generator or, if none is available, the switch's CPU.

%% file: sections/stateful.tex
\section{Operating on Persistent State}
\label{sec:stateful}

There are two kinds of state in a data-plane program: local state that lives for the processing of a single packet and global (or persistent) state that remains across packets. 
In prior data-plane languages, persistent state makes development unreasonably challenging. The problem is not that the hardware has constraints, 
but rather that the constraints are left \emph{implicit} by the language.



For example, P4 programs store persistent state in \code{RegisterArrays} and use \code{RegisterActions} with arbitrary blocks of C-like code to operate on those arrays. It is easy to write a \code{RegisterAction} that is too complex for the underlying hardware to support, but often difficult to figure out why. The ``decision'' that a particular \code{RegisterAction} is too complex is made by part of the compiler's back-end that is far removed from the source code, for example, a target-specific assembler. When a problem occurs at this late stage, neither the programmer nor the compiler have any direct way of figuring out what went wrong. For example, here is a \code{RegisterAction} body that is too complex for the Tofino and results in an assembler error related to operand referencing.
\begin{lstlisting}[style=P4Block]
void apply(inout bit<32> memCell) {
    if (memCell > y) { 
        memCell = memCell + y;
    } else {
        memCell = x + y; 
    }
}
\end{lstlisting}



\lang's solution is a carefully designed interface to persistent state that enables syntactic checks on untransformed source code. These checks occur at the very beginning of compilation, quickly identify invalid programs, and return \emph{source-level} error message that tell us exactly what is wrong.





\subsection{The Array Module}


\lang programs store persistent state in arrays, such as \inl{pathlens} in this example from the \exprog. 
\begin{lstlisting}[style=DptBlock]
global pathlens  = new Array<<32>>(tbl_sz);
memop incr(int stored, int added) { return stored + added; }
fun int get_pathlen(int dst) {
    return Array.get(pathlens, dst, incr, x);
}
\end{lstlisting}
All computation on arrays is done through \lang's \inl{Array} module, whose methods abstract the stateful operations that are possible within a single ALU. In the above example, the \inl{Array.get} method retrieves \inl{pathlens[dst]}. \inl{Array} also includes \inl{set} and \inl{update} (i.e., set and get in parallel) methods. 

Of course, the stateful ALUs of a PISA switch can do more than read or write values from persistent memory. They can read the state, perform a small amount of computation, and then write the state back to memory and/or packet metadata~\cite{domino}. \lang's \inl{Array} module has a functional interface to these capabilities. In the above example, the third argument of the call to \inl{Array.get} is \inl{incr}, a function. \inl{Array.get} will return \inl{incr(pathlens[dst], x)}. 



\lang's interface to state is flexible and allows programmers write modular and re-usable code. Functions like \inl{incr} can be re-used in any call to \inl{Array.get}, \inl{set}, or \inl{update}. Further, arrays themselves can be arguments to functions or handlers. 


\subsection{Memop Functions}
Function arguments to \inl{Array} methods, such as \inl{incr} in the above example, are \emph{memops}: a special kind of function that is syntactically restricted to ensure that it does not do more computation than what a single stateful ALU can support. 
\revised{
The syntax of a \emph{memop} is limited by the instruction set of the targeted PISA processor  because every \inl{Array} method that uses a \emph{memop} must be able to compile to a valid instruction. At the same time, the syntax must carefully balance \emph{expressiveness} and \emph{regularity}. On the one hand, we would like \emph{memops} to be flexible enough to implement any program that the underlying hardware can support. On the other hand, we would like \emph{memops} to be as simple and regular as possible, to decrease the Lucid learning curve
make it easier to use. 

With those design criteria in mind, we defined a \emph{memop} to be a function of two 
arguments that satisfies the following constraints:
\begin{itemize}[leftmargin=*]
    \item the body is either a single return statement or an if statement containing one return statement in each branch; 
    \item each variable is used at most once per expression; and
    \item only ALU-supported operators are used.
\end{itemize}
When a user declares a \emph{memop}, we automatically check that its body satisfies these requirements. If this check passes, the programmer is guaranteed that the operation is compilable; if not, our compiler can explain exactly what is wrong. 



\rerevised{
The current \emph{memop} syntax slightly favors regularity over expressiveness. There are expressions that the Tofino can implement, but \emph{memops} disallow including compound conditional expressions, operations that read multiple local variables, and approximate exponentiation. We disallow the above expressions because they can only be used in certain cases. For example, an \inl{Array.set} call that uses a \emph{memop} with a compound condition, \EG, \inl{((x == 1) || (x == 2))}, can compile to a legal sALU instruction. However, an \inl{Array.update} call that uses two \emph{memops}, each with a different compound condition, cannot compile to a legal sALU. Alternate kinds of \emph{memops} could support these expressions where possible, for example, a \emph{memop} with compound conditions that can be used in \inl{Array.set}, but not \inl{Array.update}. So far, however, we have been able to write a wide range of applications (see Figure~\ref{tab:usecaseeval}) without needing to introduce this extra complexity into the programming model. Appendix~\ref{app:memop} discusses \emph{memop} limitations further.
}




Although our current memop definition is geared specifically for the Tofino, the
principle the design suggests is quite general:  Use static, source-level constraints
to limit the expressions programmers write, as they write them.  Doing so makes it possible
to provide targeted programmer feedback that pinpoints the exact line and character where
an error occurred and ultimately saves one of the most important resources, programmer
time.
}


%% file: sections/type.tex
\section{Ordered Data Access}
\label{sec:type}



Most data-plane programs with integrated control use \emph{multiple} persistent variables. For example, the \exprog has a next hop array and link status array. 
A general constraint of any PISA processor is that such persistent data must be partitioned across the stages of a feed-forward pipeline. This leads to a natural order in which a program can access the data. Current languages force programmers to manually track the order of data access in their programs, which compounds the state-related challenges described in Section~\ref{sec:stateful}.

\begin{figure}[t]
\begin{lstlisting}[style=DptBlock]
    const int SIZE = 16;
    
    global arr1 = new Array<<32>>(SIZE);
    global arr2 = new Array<<32>>(SIZE);
    
    handle setArr1(int idx, int data) {
        int x = Array.get(arr2, idx);
        Array.set(arr1, idx, x);
    }
    handle setArr2(int idx, int data) {
        int x = Array.get(arr1, idx);
        Array.set(arr2, idx, x);
    }
\end{lstlisting}
\vspace{-4mm}
\caption{A disordered program.}
\vspace{-4mm}
\label{fig:badordering}
\end{figure}

To illustrate this issue, Figure~\ref{fig:badordering} presents a simple but invalid \lang program. The program declares two arrays \inl{arr1} and \inl{arr2}, and two handlers \inl{setArr1} and \inl{setArr2} which access those arrays in different orders. In general, programs of this form cannot be compiled to a PISA pipeline---one handler demands that data for \inl{arr1} appear earlier in the pipeline than data for \inl{arr2} and vice versa, creating irresolvable constraints. 

These constraints are fundamental to any PISA pipeline, but they are not enforced by P4. If we write the program from Figure~\ref{fig:badordering} in P4 for the Tofino, compilation does not fail until the Tofino backend. When the backend cannot solve the program's layout constraints to allocate stateful data to particular stages of the pipeline, it fails with an error that states ``Table placement cannot make any more progress'', but does not indicate what is wrong with the program. 

\lang resolves this issue by interpreting a program's data declarations as an implicit, high-level specification of the programmer's data layout intentions.  The \lang type system then verifies that the order of data accesses in the rest of the program is consistent with the specification, guaranteeing that compilation is possible (if enough pipeline stages are available).  When an access ordering error arises, a useful \emph{source-level} error message indicates the specific lines of code in conflict.


\subsection{Well-ordered Programs}

We say a \lang program is \emph{well-ordered} if the data \emph{accesses} in every handler follow the same order as the global data \emph{declarations}.  In other words, we treat the order of data declarations as a \emph{specification} that clearly documents requirements for all handlers.  It is an easy specification for programmers to write, as they must declare and initialize their data anyway.

The specification is high level---it does not refer to specific hardware stages;  in fact, programs such as this are portable across hardware platforms with different numbers of stages. The compiler has the flexibility to place any object in any stage so long as it faithfully implements the program's semantics.  The specification merely ensures that, provided a program adheres to its requirements, the compiler can find \emph{some} solution to the data-allocation problem.  

If programmers do not adhere to the specification, a simple error message directs them to the disordered portion of their program. For the program in Figure~\ref{fig:badordering}, \lang would issue an error for the \inl{setArr1} handler saying that it accesses \inl{arr2} and \inl{arr1} in the opposite order of their declarations. 
While verifying a toy example like Figure~\ref{fig:badordering} is straightforward, the verification problem becomes increasingly difficult as programs grow and use auxiliary functions to encapsulate common idioms and user-defined abstractions. Such functions may access global variables (directly or via arguments), which constrains the order in which they may be called from other functions or handlers. 

\subsection{Types for Ordered Data Access}\label{sec:types}
We use a \emph{type-and-effect} system to check that a \lang program is well-ordered while also performing regular type checking. \revised{The full details of this type system appear in Appendix~\ref{app:types}.}

While past work~\cite{vault,igarashi+:resource-usage-analysis} explored the use
of ordering constraints to ensure correct access to volatile state in other contexts (\emph{e.g.}, ``no-use-after-free'' properties for memory managers or ``no data access without first acquiring a lock''), we are unaware of prior uses of ordered type systems for data layout along pipelines, or more generally in the context of networking.  On the one hand, systems such as ordered logic~\cite{polakow-ordered} appear too restrictive for our purposes---functions
that refer to an ordered variable cannot be declared until prior variables are used.  On the other hand, prior systems~\cite{vault,igarashi+:resource-usage-analysis} designed for enforcing protocols such as open-read/write-close sequences
over OS resources are unnecessarily complex, requiring additional type annotations or sophisticated inference mechanisms. In \lang, the key difference is that we know all the ordered variables in advance, allowing us to create a simpler system that can still define and verify functions separately from where they are used.

At a high level, our strategy is to use a system in which effects are integers representing abstract \emph{stages}. Each ordered variable (either an array or a counter) is associated with a stage based on the order in which variables are declared.
During typechecking, we keep track of a \emph{current stage}, which tracks the most recently-accessed ordered variable. The typechecking fails if the program attempts to access an ordered variable whose stage is less than the current one, indicating that during execution the packet would have already passed by that data in the pipeline. 

Formally, our typing judgement has the following form: \\\centerline{$\Gamma, {\epsilon}_1 \vdash e : \tau, {\epsilon}_2$.}\\ In English, this can be interpreted as the statement ``Starting with environment $\Gamma$ and at stage ${\epsilon}_1$, the expression $e$ has type $\tau$ and will finish evaluating in stage ${\epsilon}_2$''. We use this to prove the following soundness theorem:

\textbf{Theorem:} If $\emptyset, {\epsilon}_1 \vdash e : \tau, {\epsilon}_2$, then either $e$ is a value, or $e \rightarrow e^{\prime}$ and there is some ${\epsilon}_1^{\prime}$ such that $\emptyset, {\epsilon}_1^{\prime} \vdash e^{\prime} : \tau, {\epsilon}_2$.

This theorem implies that any program which typechecks will never "get stuck" trying to access unavailable data. The 
proof of the theorem appears in Appendix \ref{app:types}.

%% file: sections/compiling.tex
\begin{figure}[t]
\begin{lstlisting}[style=DptBlock]
Array nexthops = new Array<<32>>(NUM_HOSTS);
Array pcts = new Array<<32>>(NUM_PORTS_X3);
Array hcts = new Array<<32>>(NUM_HOSTS);
memop plus(int cur, int x){return cur + x;}

event count_pkt(int dst, int proto);
handle count_pkt (int dst, int proto) {
    int idx = Array.get(nexthops, dst);
    if (proto != TCP) {
        if (proto == UDP)
            idx = idx + NUM_PORTS;
        else 
            idx = idx + NUM_PORTS_X2;
    }
    Array.set(pcts, idx, plus, 1);      
    if (proto == TCP)
        Array.set(hcts, dst, plus, 1);
}
\end{lstlisting}
\includegraphics[width=.9\linewidth]{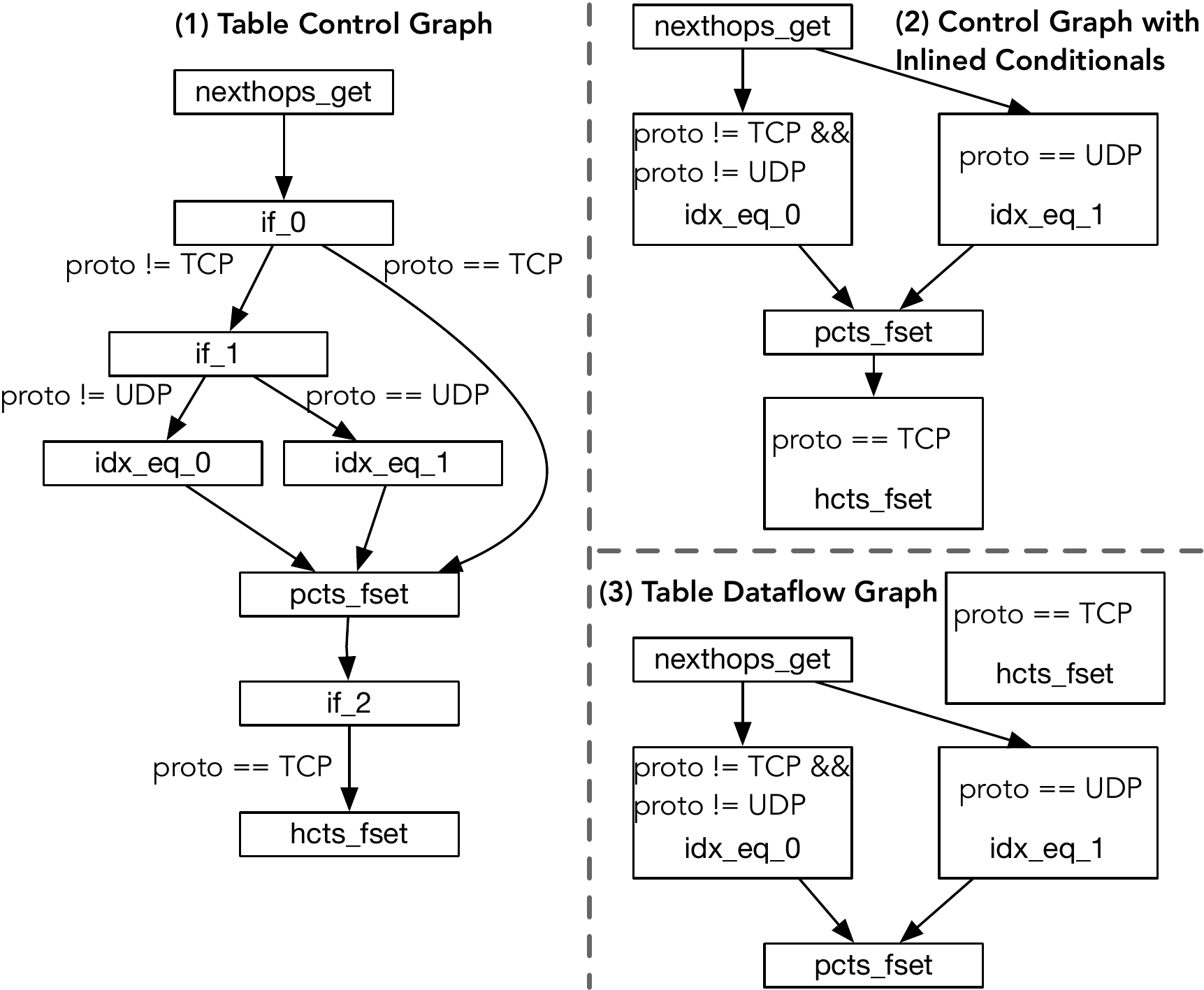}
\vspace{-4mm}
\caption {Top: a \lang handler using only atomic statements. 
Bottom: the handler represented as an atomic table graph (1) and  
optimized to require fewer pipeline stages (2 and 3).}
\vspace{-4mm}
\label{fig:cttblgraph} 
\end{figure}

\newsavebox\opbox
\begin{lrbox}{\opbox}
\begin{lstlisting}[style=SmallP4Snip]
action do_idx_add {idx = idx + NUM_PORTS;}
table tbl_idx_add { 
  actions = {do_idx_eq;}
  const default_action = {do_idx_eq;}
}
\end{lstlisting}
\end{lrbox}

\newsavebox\memopbox
\begin{lrbox}{\memopbox}
\begin{lstlisting}[style=SmallP4Snip]
RegisterAction<...>(tcp_cts) setm_1 = {
  void apply(inout bit<32>m, out bit<32>r){
    mem = mem + 1;
  } 
};
action do_setm_1() { setm_1.execute(port);}
table tbl_setm_1 { 
  actions = {do_setm_1;}
  const default_action = {do_setm_1;}
}
\end{lstlisting}
\end{lrbox}

\newsavebox\branchbox
\begin{lrbox}{\branchbox}
\begin{lstlisting}[style=SmallP4Snip]
action if_true(); action if_false();
table tbl_if {  
  keys = {proto : ternary;}
  actions = {if_true; if_false;}
  entries = {
    (TCP) : if_false;
    (_)   : if_true;
  }
}
\end{lstlisting}
\end{lrbox}

\begin{figure*}[!t]
\centering
\footnotesize
\setlength{\tabcolsep}{2pt}
\begin{tabularx}{\linewidth}{ l >{\hsize=.8\hsize}X X >{\hsize=.6\hsize}X } 
    \toprule
    \textbf{Lucid} & \inl{idx = idx + NUM_PORTS;} & \inl{Array.setm(tcp_cts, port, plus, 1);}  & \inl{if(proto != TCP)} \\
    \midrule
    \textbf{P4} & \textbf{Operation Table} & \textbf{Memory Operation Table} & \textbf{Branch Table} \\
    & 
    \usebox\opbox & \usebox\memopbox & \usebox\branchbox \\
    \arrayrulecolor{black}\bottomrule
\end{tabularx}
\vspace{-4mm}
\caption{Examples from Figure~\ref{fig:cttblgraph} of the three kinds of Atomic P4 tables \vspace{-3mm}
that the \lang compiler generates.}
\label{fig:atomictbls}
\end{figure*}

\section{Compiling to the Tofino}
\label{sec:compiler}
After syntax and type checking, the \lang compiler translates a program into \texttt{P4_16} optimized for the Intel Tofino. Most of the backend's complexity lies in compiling handler bodies, since events map directly to packet headers and the event scheduler (Section~\ref{ssec:sched}) is mostly static code. The main steps of handler compilation are translating to atomic P4 tables and optimizing control flow.

\subsection{Translating to Atomic P4 Tables}

The compiler first uses function inlining and subexpression elimination to reduce a handler's body into a graph of statements that are each simple enough to execute with at most one Tofino ALU. The top of Figure~\ref{fig:cttblgraph} is an example of a handler where all statements are already in this atomic form. The compiler then translates each atomic statement directly into a P4 table, to produce an atomic table graph like the one shown in Figure~\ref{fig:cttblgraph}(1). There are three kinds of atomic tables, demonstrated in Figure~\ref{fig:atomictbls}.

\topheading{Operation table.} An operation table uses a single ALU to evaluate a binary expression over two local variables (metadata in P4) and assign its result to a third local variable. 

\topheading{Memory operation table.} A memory operation table uses a single stateful ALU to update one element in a P4 register array. It is a direct translation of a call to an Array method. 


\topheading{Branch table.} A branch table uses two match-action rules to compare a local variable against a constant to determine which table will execute next. 


\subsection{Optimizing Control Flow}
\label{ssec:p4optimizations}

The \lang compiler optimizes the atomic table graph in three steps to reduce the number of pipeline stages it requires. 

\heading{Inlining branch operations.}
Branch tables are wasteful in the table representation of a \lang program because, in a PISA pipeline, a branch table's successors must be placed in a subsequent stage. 

The compiler eliminates branch tables by first transforming each non-branch table to check the conditions necessary for its own execution using static match-action rules. The table executes its single action if there is a match, else a no-op. For example, by following the path from the root of Figure~\ref{fig:cttblgraph}(1) to the table \inl{idx_eq_0}, we see that it only executes if \inl{(proto != TCP) && (proto != UDP)}, thus in Figure~\ref{fig:cttblgraph}(2), the table \inl{idx_eq_0} tests these conditions before executing. 

The compiler applies this transformation to all tables then deletes the branch tables. As Figure~\ref{fig:cttblgraph}(1) and (2) show this saves three pipeline stages in the example program.

\heading{Rearranging tables.}
Next, the compiler rearranges tables based on data flow to further reduce the number of stages used by a program. For example, in Figure~\ref{fig:cttblgraph}(2), the table that implements \inl{Array.set(hcts, dst, plus, 1);} (\inl{hcts_fset}) does not have any data flow dependencies on previous tables. None of the variables that it reads are modified by any other tables in the program, so it can be executed in parallel with the first table, as shown in Figure~\ref{fig:cttblgraph}(3). 


\begin{figure}[t]
    \includegraphics[width=.9\linewidth]{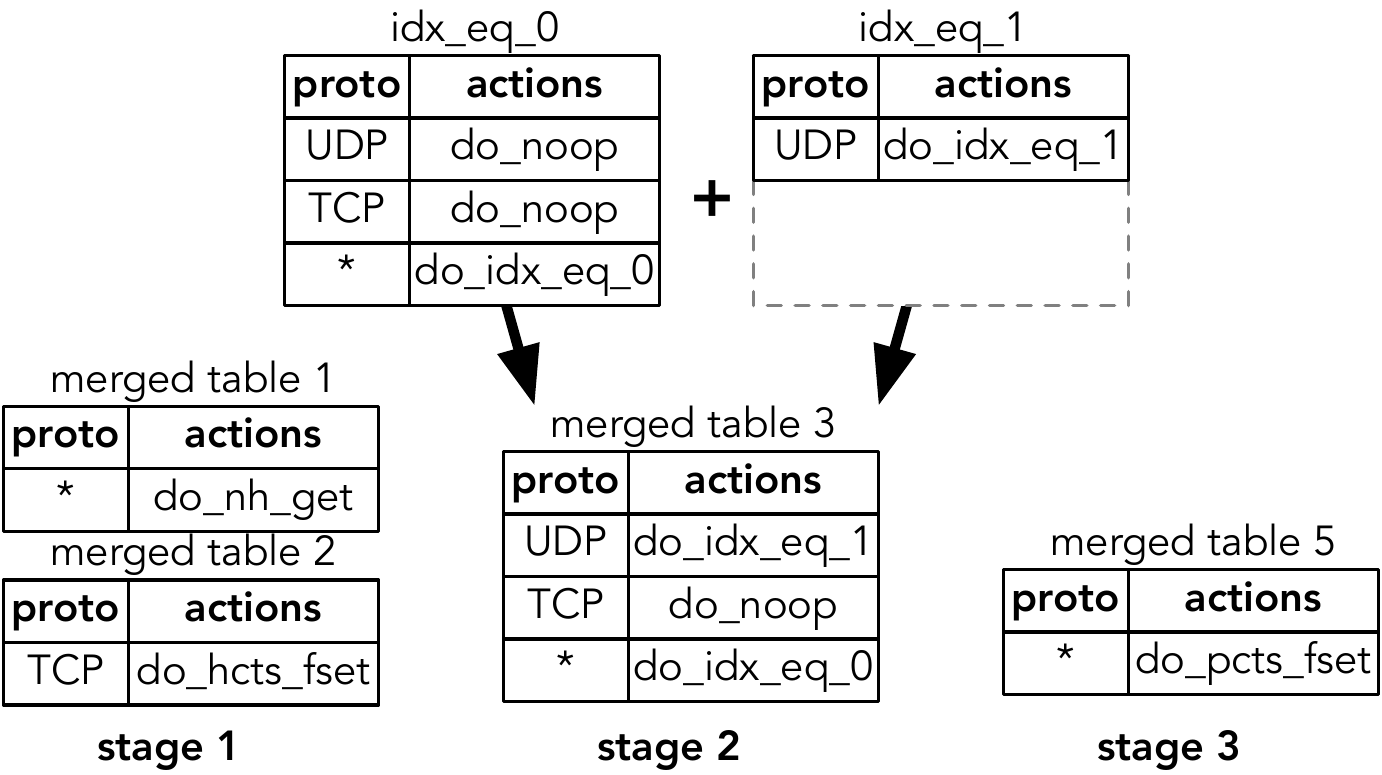}
\vspace{-4mm}
\caption {Merged tables for the program in Figure~\ref{fig:cttblgraph}.}
\vspace{-3mm}
\label{fig:mergedtbls} 
\end{figure}

\heading{Merging tables and actions.}
For complicated programs, the atomic table representation of a \lang program would still require many stages because it uses many tables (one per operation) and PISA processors can only support a limited number of tables per stage. \lang's final optimization reduces table overhead by merging atomic tables into multiple-operation tables. This is possible because \lang-generated tables use only static rules. Figure~\ref{fig:mergedtbls} shows how tables from the example in Figure~\ref{fig:cttblgraph} get merged. 

The compiler uses a simple greedy algorithm that produces a pipeline with $M$ stages and $N$ merged tables per stage by walking the atomic table graph topologically. For each table $t$, it finds the earliest merged table that $t$ can be merged into. This decision is based on data flow constraints, a simple model of the free resources in each stage, and a small number of Tofino-specific constraints.

When a merged table $m$ that can fit $t$ is found, $m$ is replaced with $m'$---the cross product of $m$ and $t$. The algorithm ends when either all atomic tables have been merged, or it reaches an atomic table that cannot be placed. 

%% file: sections/eval.tex
\section{Evaluation}
\label{sec:eval}

\begin{figure}[!t]
\centering
\footnotesize
\setlength{\tabcolsep}{2pt}
\begin{tabularx}{\linewidth}{ l X c c c c c } 
    \toprule
 & & \multicolumn{2}{c}{\textbf{LoC}} & {\textbf{Tofino}} \\
\textbf{Application} & \textbf{Description} & \textbf{\lang} & \textbf{P4} &\textbf{Stages} \\

\midrule
	\makecell[tl]{Stateful\\Firewall\\(SFW)} & \fnwords{Blocks connections not initiated by trusted hosts. \emph{\textbf{ Control events update a Cuckoo hash table.}} } & 189 & 2267 & 10   \\
    \arrayrulecolor{black!50}\midrule
	\makecell[tl]{Fast\\Rerouter\\(RR)} & \fnwords{Forwards packets, identifies failures, and routes. \emph{\textbf{Control events perform fault detection and routing.}}} & 115 & 899  & 8  \\
    \arrayrulecolor{black!50}\midrule
	\makecell[tl]{Closed-loop\\DNS Defense\\(DNS)} & \fnwords{Detects/blocks DNS reflection attack with sketches \& Bloom filters. \emph{\textbf{Control events age data structures.}} }& 215 & 1874 & 10 \\
    \arrayrulecolor{black!50}\midrule
	\makecell[tl]{*Flow~\cite{starflow}} & \fnwords{Batches packet tuples by flow to accelerate analytics. \emph{\textbf{Control events allocate memory.}}} & 149 & 1927 & 12 \\
    \arrayrulecolor{black!50}\midrule
	\makecell[tl]{Consistent\\Shared State\\(SRO)\cite{10.1145/3422604.3425946}} & \fnwords{Strongly consistent distributed arrays. \emph{\textbf{Control events synchronize writes.}}} & 94 & 897 & 11 \\
    \arrayrulecolor{black!50}\midrule
	\makecell[tl]{Distributed\\Prob. Firewall\\(DFW)} & \fnwords{Distributed Bloom filter firewall. \emph{\textbf{Control events sync. updates.}}} & 66 & 1073 & 10 \\
	\makecell[tr]{+Aging} & \fnwords{\emph{\textbf{Adds control events for aging.}}} & 119 & 1595 & 10 \\
    \arrayrulecolor{black!50}\midrule
	\makecell[tl]{Single-dest.\\RIP} & \fnwords{Routing with the classic Route Information Protocol (RIP). \emph{\textbf{Control events distribute routes.}}} & 81 & 764 & 8\\
    \arrayrulecolor{black!50}\midrule
	\makecell[tl]{Simple NAT} & \fnwords{Basic network address translation. \emph{\textbf{Control events buffer packets and install entries.}}} & 41 & 707 & 11\\
    \arrayrulecolor{black!50}\midrule
	\makecell[tl]{Historical\\Prob. Queries\\(CM)} & \fnwords{Measures flows with sketches for historical queries. \emph{\textbf{Control events age and export state periodically.}}}  & 93 & 856 & 5 \\
	\arrayrulecolor{black}\bottomrule
\end{tabularx}



\vspace{-4mm}
\caption{Applications with data plane-integrated control, implemented in \lang and compiled to the Barefoot Tofino. The role of control events is bolded.}
\vspace{-4mm}
\label{tab:usecaseeval}
\end{figure}


We evaluate \lang by implementing the applications described in Figure~\ref{tab:usecaseeval} and compiling them to the Tofino with the \lang compiler~\footnote{\url{https://github.com/princetonUniversity/lucid}} and P4-studio verision 9.2. We analyze the design of \lang, the effectiveness of its optimizations and the runtime overhead of recirculation. Finally, a case study with the stateful firewall evaluates the potential performance benefit of data-plane integrated control. 


\subsection{Language Design}

We first compare the lines of code required to program in \lang versus P4. *Flow~\cite{starflow} is a complex application that gives us a point of comparison to hand-written P4. The \lang program is a complete implementation of *Flow in 149 lines of code. The published implementation, in P4\_14, is 1559 lines of code---over 10X longer.


We are unaware of hand-written implementations for the other \lang applications and, due to the time required to program the Tofino in P4, we did not re-implement any ourselves. However, using *Flow as a calibration point suggests that the \lang compiler produces P4 that is within ~15\% the length of hand-written P4. Thus, we conclude that \lang reduces lines of code by around 10X for diverse applications.

\begin{figure}[t]
\includegraphics[width=.9\linewidth]{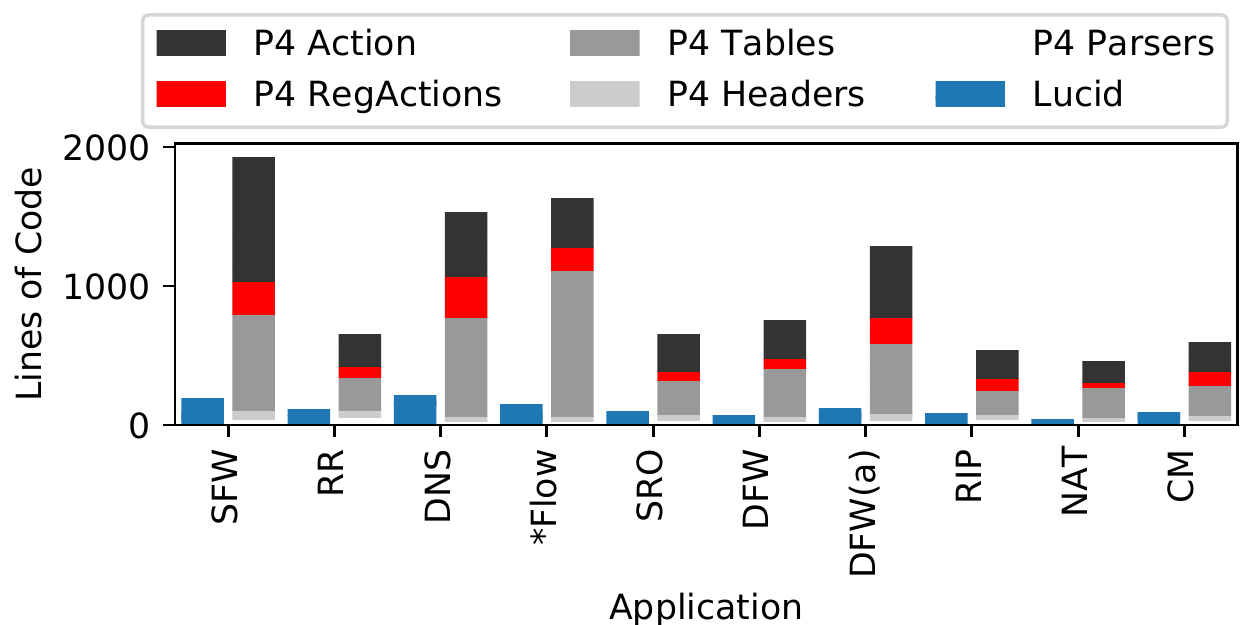}
\vspace{-4mm}
\caption {Breakdown of P4 code in Figure~\ref{tab:usecaseeval}.}
\vspace{-2mm}
\label{fig:loc_breakdown} 
\end{figure}

Figure~\ref{fig:loc_breakdown} breaks down the lines of P4. Actions and tables take the most lines. An interesting observation is that for most applications, the entire \lang program was fewer lines of code than just the register actions in P4. This is partially because P4 register actions are not reusable like \lang's memops---the programmer must manually copy the code every time they want to repeat the same operation on a different array. In \lang programs, and even across programs, we re-use the same generic memops multiple times.

\begin{figure}[!t]
\centering
\small
\setlength{\tabcolsep}{4pt}
\begin{tabularx}{\linewidth}{ l c c c c } 
    \toprule
	\textbf{Application} 			& NAT 			& {RIP} 	   & {Dist FW} 		  & {Dist FW + Aging} \\
	\textbf{Dev. Time}				& 25m			& 40m 		   & 25m			  & 25m + 30m 		 \\
	\bottomrule
\end{tabularx}
\vspace{-4mm}
\caption{Time for a student without Tofino experience to write Tofino-compiling \lang applications.}
\vspace{-4mm}
\label{tab:impltimes}
\end{figure}

\revised{
    The applications in Figure~\ref{tab:impltimes} were implemented by a PhD 
    student who has never programmed the Tofino before. As it shows, it took less than an hour to write each of these non-trivial prototypes that successfully compiled to the Tofino. With P4, this level of productivity is hard to imagine, even for an experienced Tofino programmer. For PhD students new to the architecture, it can take weeks to do anything non-trivial. \rerevised{We are excited by the potential for \lang to save future students' time.} 
}



\subsection{Optimization Benchmarks}

\begin{figure}[t]
\centering
\begin{minipage}{.49\linewidth}
\includegraphics[width=.9\linewidth]{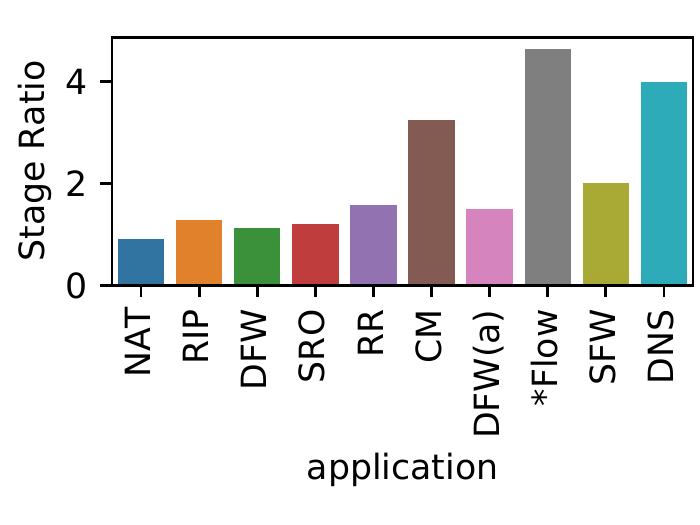}
\vspace{-4mm}
\caption {Optimized stage count vs unoptimized.}
\vspace{-4mm}
\label{fig:cp_compression} 
\end{minipage}
\hfill
\begin{minipage}{.49\linewidth}
\includegraphics[width=.9\linewidth]{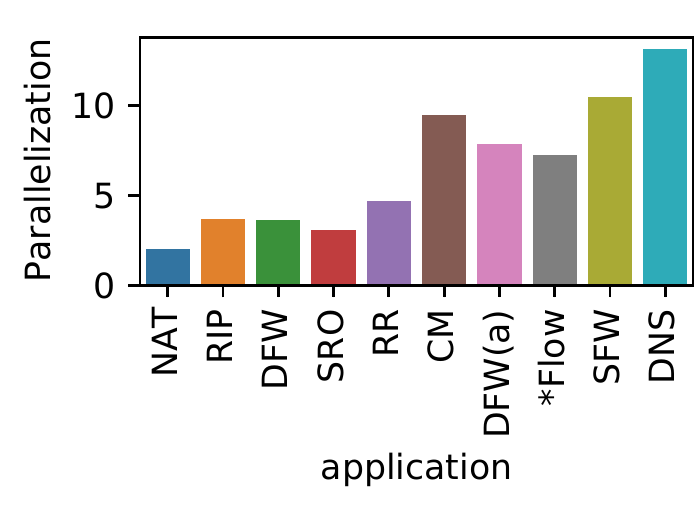}
\vspace{-4mm}
\caption {ALU instrs. per stage in optimized code.}
\vspace{-4mm}
\label{fig:instrs_per_stage} 
\end{minipage}
\end{figure}



\heading{Compiler optimizations.}
Next, we evaluate \lang's compiler optimizations by comparing the number of required stages with and without optimizations. Figure~\ref{fig:cp_compression} shows the ratio for each application. For unoptimized stages, we report the number of atomic P4 tables in the longest code path, as many programs did not fit into the Tofino's pipeline without optimization. Optimizations reduced stage requirements by a factor of 1.5-4 for most applications. The benefit was greater for complex applications, such as *Flow and the closed-loop DNS defense system, which originally had critical paths nearly 4X too long for the Tofino's pipeline.

Figure~\ref{fig:instrs_per_stage} shows the number of \lang statements that the compiler mapped to each stage. It ranged from 2 - 13, demonstrating that the compiler was able to find and exploit a significant amount of parallelism in the programs. 

\begin{figure}[t]
\centering
\begin{minipage}{.49\linewidth}
\includegraphics[width=\linewidth]{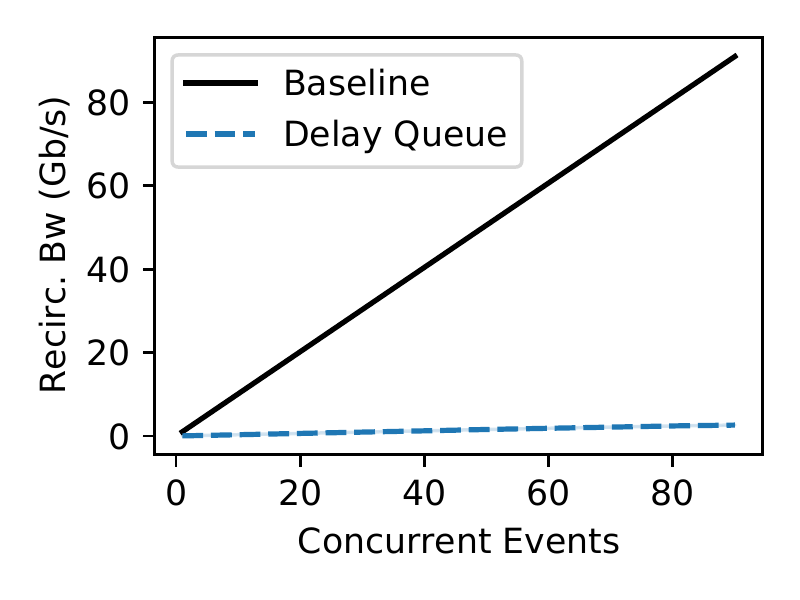}
\end{minipage}
\hfill
\begin{minipage}{.49\linewidth}
\includegraphics[width=\linewidth]{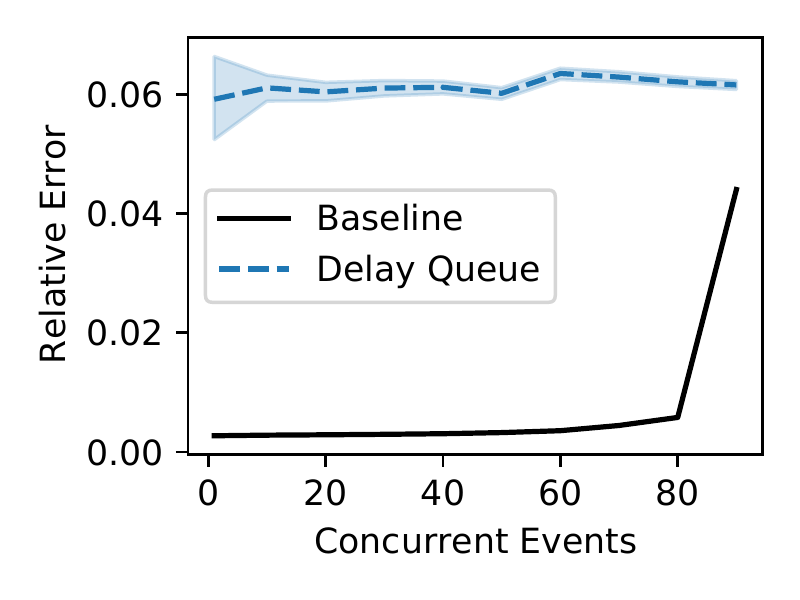}
\end{minipage}
\vspace{-4mm}
\caption {Pausable queue overhead and accuracy.}
\vspace{-4mm}
\label{fig:pause_queue_bench} 
\end{figure}

\heading{Event scheduler optimizations.} 
A key optimization in \lang's event scheduler is the pauseable queue mechanisms for reducing the overhead of delaying events via recirculation. We measured the bandwidth overhead and timing accuracy of delaying 64B event packets on one of the Tofino's 100 Gb/s recirculation ports, with and without the pausable queues.

As Figure~\ref{fig:pause_queue_bench} shows, the pausable queues make the recirculation bandwidth cost of delayed events negligible. The bandwidth cost for delaying 90 concurrent events indefinitely was 5.5 Gb/s. In comparison, delaying 90 concurrent events \emph{without} \lang's pausable queues consumed over 95 Gb/s---the port was effectively saturated.

\revised{
This nearly 20X reduction in overhead has two costs: increased packet buffer utilization and timing variance. The increase in packet buffer utilization is small  compared to the amount of recirculation throughput saved. For example, storing 90 64B events in a queue uses around 7KB of packet buffer (depending on memory cell size). The Tofino has 22MB of shared packet-buffer memory, or a bit more than 320KB per port. So, with 90 concurrent events we trade around \emph{2\%} of the recirculation port's fair share of buffer space for a 20X reduction in bandwidth utilization.
}

Pausable queues also increases the variance of event execution times. As Figure~\ref{fig:pause_queue_bench} shows, event delay was off by up to approximately \us{50} when using pausable queues that release every \us{100}.


\revised{
\subsection{Recirculation Overhead}
\label{ssec:recircoverhead}
\begin{figure}[!t]
\rerevised{
\centering
\small
\setlength{\tabcolsep}{4pt}

\begin{tabularx}{\linewidth}{ X c X } 
    \toprule
        \textbf{Recirc. use} & \textbf{Recirc. rate} & \textbf{Applications} \\
 {Data struct. maintenance}  & O($\frac{\text{num. entries}}{\text{scan interval}}$) & \fnwords{SFW, RR, DWF, CM, DNS, RIP} \\ 
    {Flow setup} & E[O($\text{flow rate}$)] & \fnwords{SFW, NAT, *Flow, RR} \\
   {State synchronization} & O($\text{update rate}$) & \fnwords{SRO, DFW} \\

    \bottomrule    
\end{tabularx}
\vspace{-4mm}
\caption{Recirculation uses in Figure~\ref{tab:usecaseeval} applications.}
\vspace{-3mm}
\label{tab:recircuses}
}
\end{figure}

\rerevised{
As Figure~\ref{tab:recircuses} shows, the example applications recirculate packets to perform: data structure maintenance, in which case a timed loop triggers periodic recirculation; flow setup, in which case new flows trigger recirculation; or state synchronization, in which case a state update event must recirculate through multiple switches. The stateful firewall requires the most recirculation out of all of the applications. It features both a flow setup operation, installing per-flow entries into a Cuckoo hash table, and a data structure maintenance operation, scanning the Cuckoo hash table to find and delete timed-out flows.
}


To understand recirculation overhead more concretely, we analyze the stateful firewall in more detail. We base analysis on an idealized PISA processor with a throughput of 1B packets per second that services 10 100Gb/s front-panel ports plus a 100Gb/s recirculation port. This processor supports line rate on all front-panel ports simultaneously when packets are larger than 125B and the recirculation port has no load.

Given this idealized platform, we derive a simple explanatory model of the stateful firewall's recirculation rate. Model parameters are: $N$, the size of the firewall's table; $i$, the per-flow timeout check interval; and $f$, the flow-arrival rate. The worst-case recirculation rate, $r$, is: $r=\frac{\text{N}}{\text{i}} + f\cdot\log(N)$. The first term is recirculation for timeout scanning and the second term is \emph{worst-case} recirculation for flow installation, as an installation in a Cuckoo table may require $\log(N)$ Cuckoo operations, each of which requires a recirculation.


\begin{figure}[!t]
\centering
\small
\setlength{\tabcolsep}{4pt}
\begin{tabularx}{\linewidth}{ l c c c c c } 
    \toprule
    \textbf{flow rate} ($f$)            & 10K flows/s            & 100K flows/s   & 1M flows/s \\
    \midrule
    \textbf{recirc. rate}                & 815K pkts/s            & 2M pkts/s   & 16M pkts/s \\
    \textbf{pipeline utilization}          & 0.08\%            & 0.22\%   & 1.66\% \\
    \textbf{min. pkt. size}         & 125.26            & 125.55   & 127.67 \\

    \bottomrule
\end{tabularx}
\vspace{-4mm}
\caption{Modeled worst-case recirculation overhead for the stateful firewall with $N=2^{16}$ and $i=$\ms{100}.}
\vspace{-4mm}
\label{fig:sfwrecirc}
\end{figure}

Figure~\ref{fig:sfwrecirc} shows that the recirculation rate is high in absolute numbers, but only a small percentage of the pipeline's packet-processing bandwidth---a workload with 1M new flows per second has less than a 2\% bandwidth overhead. At this point, the pipeline could still support line rate on all front-panel ports if all packets were larger than 128B (versus 125B with no recirculation load). 

}

\subsection{Stateful Firewall Case Study}
\label{ssec:sfw}
\begin{figure}[t]
\includegraphics[width=.75\linewidth]{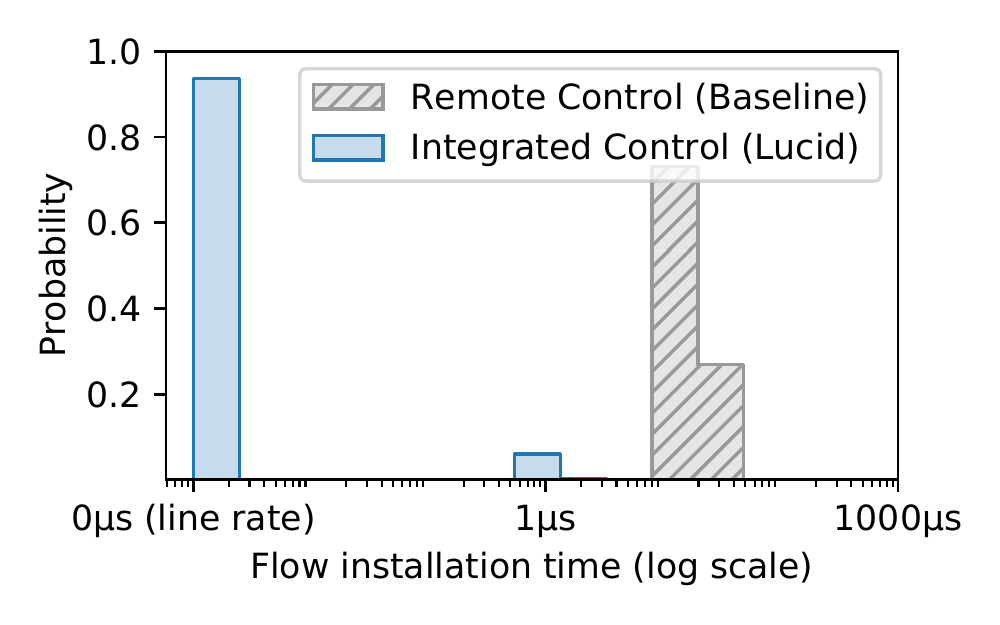}
\vspace{-6mm}
\caption {Stateful firewall flow installation times. 1000 trials using a 2048-element table with a load factor of $.3125$.}
\vspace{-6mm}

\label{fig:sfw_app_performance} 
\end{figure}
Finally, to evaluate the latency reduction that data-plane integrated control can provide, we benchmark a stateful firewall in \lang. 


\heading{Implementation.}
The core of the \lang stateful firewall is a Cuckoo hash table~\cite{pagh2004cuckoo}. Each flow maps to one of two possible locations, addressed by a different hash of its key. A lookup operation simply checks both locations in sequence. An install operation for flow $f$ attempts to place it in either of $f$'s locations. If both are occupied, a Cuckoo operation replaces a colliding victim $v$ with $f$, then generates an event to re-install $v$. This process repeats until an install operation has no victim, or the install handler detects that it has attempted to re-install $f$ more than twice, indicating failure~\cite{cuckoostash}. While a flow $v$ is being re-installed, its entry is stored in a stash at the end of the pipeline, so 
that the (re-)install operations are transparent to concurrent lookups.

\heading{Flow installation time.}
Our benchmark metric is flow installation time---the difference between when the first packet of a flow arrives and when the corresponding installation operation completes. This metric is critical in a stateful firewall because flow installation must complete in between the arrival of the first packet from an outbound flow and the arrival of the first packet from the return flow, \IE, one RTT. If not, the firewall will disrupt traffic by either dropping return packets or queuing the first packet of every new flow until the installation is complete.

\heading{Baseline.} 
We compare against a baseline that represents flow installation with efficient remote control, using Mantis~\cite{mantis}. Mantis is a driver-level framework for low-latency control in the management CPU of a Tofino switch. We measure the time required for a Mantis control thread to install a new entry into a P4 match-action table in the Tofino. This is a lower bound because it ignores the time required for the CPU to detect that a new flow has arrived, for example by polling a P4 register in the Tofino that stores a ring buffer of new flow keys.



\heading{Benchmarks.} Figure~\ref{fig:sfw_app_performance} shows the distribution of flow installation times. Average flow installation time for the data-plane integrated version was only \ns{49}. For over 90\% of flows, installation completed during the processing of the flow's first packet---an effective flow installation time of \ns{0}. Most of the remaining flows installed in a single recirculation---about \ns{600}. The worst case was 4 recirculations or around \us{2.4}. In comparison, the remote-controlled baseline took at least \us{12} to install a rule for a new flow into a P4 match-action table, with an average of \us{17.5}---\emph{over 300X longer than the data-plane integrated version}. End-to-end flow installation time with remote control would be much higher in practice, because of the time required to inform the remote controller that a new flow has arrived and the queuing delays that would occur when multiple flows arrived simultaneously. 

\revised{
\heading{Memory efficiency.} A drawback of the current \lang firewall prototype is memory efficiency. It uses around 3X more memory than the remotely controlled baseline. The \lang version does not include mechanisms to resolve or eliminate installation failures~\cite{cuckoostashprovable, cuckoostash}, so the load factor must be kept low ($0.3125$ in our experiments) to keep the probability of flow installation failure low. The remotely-controlled baseline, on the other hand, uses the native match-action tables of the Tofino that are based on a more sophisticated Cuckoo hashing algorithm that supports a load factor near $1$. Improved algorithms for Cuckoo hashing are possible in \lang, for example we implemented a Cuckoo hash table with a stash~\cite{cuckoostash}. We leave exploration of more advanced \lang data structures for future work. 
}

%% file: sections/related.tex
\revised{
\section{Discussion}


\lang's abstractions, syntactic restrictions, and backend optimizations go a long way towards making data-plane programming feel less like arcane magic and more like software engineering. Still, \lang is a work in progress. This section discusses the current limitations of \lang and data-plane integrated control.


\subsection{Language Limitations} 
\lang's limitations arise from our design goal: 
a high-level language that lets us reliably write and compile self contained applications to a specific, widely-used PISA processor.

\heading{Read-only tables.}
    \lang does not provide a direct abstraction of a PISA processor's
TCAM-based match-action tables. 
Applications can use match-action tables to classify packets in P4 that applies to packets before/after the \lang event dispatcher is called. Updating these tables is slow because it must involve the switch's CPU. For higher performance, programmers can build ``software'' packet-classification data structures~\cite{gupta2001algorithms} in \lang, like the Cuckoo hash table in the
stateful firewall. These data structures update faster, but may have recirculation overhead and require more stages compared to lower-level P4 equivalents.


\heading{Portability.}
\lang currently only targets the Intel Tofino. Portable data-plane languages are a focus of recent research~\cite{lyra}. The challenge of portability arises from the diversity of data plane hardware. P4 programs, for example, are often not portable because many commonly-used primitives (such as those for stateful operations) are platform-specific externs. 
Although \lang currently only supports the Tofino, its event-based abstractions and ordered type-and-effect system are relevant to any PISA processor. The concept of \emph{memops} is also portable, though the syntax of a \emph{memop} may change depending on target capabilities~\cite{domino}. 

\rerevised{
\heading{Multiple pipelines.} Switches often have multiple PISA pipelines, \EG, one ingress and egress pipeline per 16 ports. While the current implementation of \lang does not support state sharing across pipelines, future implementations could support multi-pipeline applications by adding pipelines as event locations and extending \lang's event scheduler. 
}

\heading{Optimization.}
The optimizations in \lang's compiler are heuristics based on our experiences with hand-coding efficient P4-Tofino programs. Recent work suggests that sophisticated optimizations based on synthesis~\cite{chipmunk} or ILP~\cite{p4all} algorithms may do better. However, in general, finding an optimal PISA layout is NP complete~\cite{vass2020compiling}.


\subsection{Integrated Control Limitations}
While data-plane integrated control can have significant benefits, it is not always the best option. We identify three factors that can make remote control more appealing. 


\heading{Compute-bound operations.} Compute-bound operations~\cite{elastictree} do not benefit as much from the reduced communication overhead that data-plane integration provides. Further, for compute-bound tasks, a server may be faster than the data-plane processor~\cite{chen2020implementing}. Of course, future packet processing architectures may shift the balance~\cite{swamy2020taurus}. 

\heading{Centralized, network-wide control.} While remote control has high overhead, an advantage is that it enables a centralized programming model. Centralization reduces programmer effort, as demonstrated by prior control-plane languages like Flowlog~\cite{flowlog}, Frenetic~\cite{frenetic}, 
NetKAT~\cite{netkat}, and McNetKAT~\cite{mcnetkat}. 
Data-plane integrated control, on the other hand, has much lower overhead but a distributed programming model. An open question is whether we can provide the abstraction of logically centralized control atop a distributed layer of data-plane integrated control, to provide a simpler programming model with low communication overhead. 

\heading{Runtime overhead.} Data-plane integrated control adds runtime overhead due to packet recirculation. This overhead is often low (Section~\ref{ssec:recircoverhead}), but is application dependent and should be considered by the programmer. Future architectures could eliminate this overhead with hardware to process control operations in parallel with packets~\cite{ibanez2019event} and periodically synchronize shared state.

}

%% file: sections/concl.tex
\section{Conclusion}

\lang makes it easy to write data-plane applications with high-performance integrated control. 
PISA switches \emph{already} have all the necessary mechanisms; however, programmers today must use them at a very low level.
%
Instead of writing packet-processing functions that always execute here and now, \lang programmers can use intuitive event-based abstractions to distribute control in both time and space. 
Complementing this is a careful correct-by-construction approach to stateful operations in the data plane that uses syntactic constraints and a sound type system to improve compiler feedback and rule out programs that are unlikely to compile.


%
We realize these ideas for the Intel Tofino, with 
an optimizing compiler that generates efficient, Tofino-compatible P4. A diverse range of 
\lang applications require require $\sim$10X fewer lines of code, compared to P4 equivalents. Programmers without \emph{any} prior Tofino experience are able to write compiling 
code in well under an hour. Finally, a \lang 
stateful firewall outperforms a remotely-controlled 
baseline by over 300X. 

Overall, \lang is general, fast, and easy to use. It will save time, enable new applications, and perhaps change the way we think about what hardware data planes can do. 

\heading{Acknowledgments.} We thank our shepherd, Brent Stephens, and the anonymous reviewers for their feedback. We also thank Mihai Budiu, Ben Pfaff, Leonid Ryzhyk, and Muhammad Shahbaz for fruitful discussions and useful feedback on this project, and Dovid Braverman for his help with developing the compiler front-end. This work is supported by NSF grants CNS-1703493 and FMitF-1837030 and DARPA grants HR0011-17-C-0047 and HR0011-20-C-0160.

%% file: sections/appendix.tex
\appendix
\clearpage
\input{sections/appendixtyping}
\input{sections/appendixmemop}
\input{sections/appendixartifact}

%% file: sections/appendixtyping.tex
\noindent{\emph{Appendices are supporting material that has not been peer-reviewed.}}
\section{Lucid's Type System}\label{app:types}
\makeunderscoreactive
\begin{figure}[h]
\footnotesize
\noindent\rule{\columnwidth}{.1em}
\setlength{\grammarparsep}{7pt plus 1pt minus 1pt}
\begin{grammar}
<\ensuremath{\epsilon} (stages)> ::= 0 | 1 | 2 | \dots

<\ensuremath{T} (base types)> ::= Unit | Int

<\ensuremath{tau} (types)> ::= \ensuremath{T} | ref (\ensuremath{T, \epsilon}) | \ensuremath{(\tau, \epsilon) \rightarrow (\tau, \epsilon)}

<\ensuremath{x} (variables)> ::= alphanumeric

<\ensuremath{g} (global variables)> ::= \ensuremath{g_0} | \dots | \ensuremath{g_{n-1}}

<\ensuremath{v} (values)> ::= () | \Z\ | \ensuremath{g} | \ensuremath{\texttt{fun}\ (x:\tau, \epsilon) \rightarrow e}

<\ensuremath{e} (expressions)> ::= \ensuremath{v} | \ensuremath{x} | \ensuremath{e+e} | let \ensuremath{x} = \ensuremath{e} in \ensuremath{e} | !\ensuremath{e} | \ensuremath{e := e} | \ensuremath{e\ e}

\end{grammar}
\vspace{-0.5em}
\noindent\rule{\columnwidth}{.1em}%
\vspace{4pt}
\caption{A toy language and type system}
\label{fig:typegrammar}
\end{figure}
\makeunderscoreletter

\makeunderscoreactive
\begin{figure}[h]
\footnotesize
\noindent\rule{\columnwidth}{.1em}
\begin{mathpar}

    \relationRule{int}{
		n \in \Z
	}{
		\Gamma, \epsilon \vdash n\ \colon \texttt{Int}, \epsilon
	}
	
	\relationRule{Unit}{
		\ 
	}{
		\Gamma, \epsilon \vdash ()\ \colon \texttt{Unit}, \epsilon
	}

	\relationRule{global variable}{
		\
	}{
		\Gamma, \epsilon \vdash g_i\ \colon \texttt{ref}(T_i, i), \epsilon
	}
	
	\relationRule{local variable}{
		\Gamma[x] = \tau
	}{
		\Gamma, \epsilon \vdash x\ \colon \tau, \epsilon
	}

	\relationRule{plus}{
		\Gamma, \epsilon \vdash e_1\ \colon \texttt{Int}, \epsilon_1\\
		\Gamma, \epsilon_1 \vdash e_2\ \colon \texttt{Int}, \epsilon_2
	}{
		\Gamma, \epsilon \vdash e_1 + e_2\ \colon \texttt{Int}, \epsilon_2
	}
	
	\relationRule{let}{
		\Gamma, \epsilon \vdash e_1\ \colon \tau_1, \epsilon_1\\
		\Gamma[x := \tau_1], \epsilon_1 \vdash e_2\ \colon \tau_2, \epsilon_2
	}{
		\Gamma, \epsilon \vdash \texttt{let } x = e_1 \texttt{ in } e_2\ \colon \tau_2, \epsilon_2
	}
	
	\relationRule{deref}{
		\Gamma, \epsilon \vdash e\ \colon \texttt{ref}(T, \epsilon_1), \epsilon_2\\
		\epsilon_2 \leq \epsilon_1
	}{
		\Gamma, \epsilon \vdash\ !e\ \colon T, \epsilon_1+1
	}

    \relationRule{update}{
	    \Gamma, \epsilon \vdash e_1\ \colon T, \epsilon_1\\
	    \Gamma, \epsilon_1 \vdash e_2\ \colon \texttt{ref}(T, \epsilon_2), \epsilon_3\\
		\epsilon_3 \leq \epsilon_2
    }{
	    \Gamma, \epsilon \vdash e_2 := e_1\ \colon \texttt{Unit}, \epsilon_2+1
    }
    
    \relationRule{abs}{
		\Gamma[x := \tau_{in}], \epsilon_{in} \vdash e\ \colon \tau_{out}, \epsilon_{out}
	}{
		\Gamma, \epsilon \vdash \texttt{fun } (x : \tau_{in}, \epsilon_{in}) \rightarrow e\ \colon (\tau_{in}, \epsilon_{in}) \rightarrow (\tau_{out}, \epsilon_{out}), \epsilon
	}
	
	\relationRule{app}{
		\Gamma, \epsilon \vdash e_1\ \colon (\tau_{in}, \epsilon_{in}) \rightarrow (\tau_{out}, \epsilon_{out}), \epsilon_1\\
		\Gamma, \epsilon_1 \vdash e_2\ \colon \tau_{in}, \epsilon_2\\
		\epsilon_2 \leq \epsilon_{in}
	}{
		\Gamma, \epsilon \vdash e_1\ e_2\ \colon \tau_{out}, \epsilon_{out}
	}
	
\end{mathpar}
\vspace{-0.5em}
\noindent\rule{\columnwidth}{.1em}%
\vspace{4pt}
\caption{Typing rules}
\label{fig:typerules}
\end{figure}
\makeunderscoreletter

This appendix presents a simplified definition of \lang's type system, as well as an operational semantics and soundness proof. To begin, figure \ref{fig:typegrammar} defines a grammar for a model ML-like language on which we will define our type-and-effect system. We present the system for this language purely for convenience; adapting the rules to \lang's C-like syntax presents no theoretical challenge.

The system is defined with respect to some predefined, ordered set of $n$ global variables $g_0$ through $g_{n-1}$, each of which has an associated \emph{base type} $T_i$. Base types are simply types which do not reference stages -- in this example, the only two base types are \texttt{Unit} and \texttt{Int}. Despite the name, global variables are treated as values in the language, not as variables. They can be thought of exactly like ref cells in OCaml, and this line of thinking inspires much of the syntax used for them in the language.

Effects in this system are called \emph{stages}, and are used to track which global variables have been used so far. Intuitively, the stage i represents the pipeline stage containing $g_i$. We begin typechecking at stage 0, and increment the stage when global variables are used. We can then ensure that the global variables are used in order by only allowing variable $g_i$ to be used if the current stage is at most $i$.

The types in this system are mostly straightforward, but note that functions now have starting and ending stages as well as input and output types, and we have a \texttt{ref} type representing the type of a global variable -- in general, the type of $g_i$ is $\texttt{ref }(T_i, i)$.

Expressions in this language are also straightforward, except that we have two operators on global variables: dereference ($!e$), which returns the current value of the variable, and update $e_1 := e_2$, which updates global variable $e_1$ to hold the value of $e_2$. Both of these operators access the global variable, and hence should never be used out-of-order. Note that these operators do not exist in \lang; instead, there are several builtin functions for performing these accesses.

\subsubsection{The Typing Judgement}
Our typing judgement has the form $\Gamma, {\epsilon}_1 \vdash e : \tau, {\epsilon}_2$, where $\Gamma$ is an environment which maps local variables to values.
In English, this judgement can be read as ``starting with environment $\Gamma$ at stage ${\epsilon}_1$, the expression $e$ has type $\tau$ and will finish evaluation in stage ${\epsilon}_2$''. The typing rules are presenting in \ref{fig:typerules}.

The most interesting rules here are the DEREF and UPDATE rules, as they are the ones which interact with stages. Each first typechecks its subexpression(s) and expects to receive a global variable $g_i$ (i.e. something with \texttt{ref} type) as its first argument. Crucially, neither rule can be applied unless the stage after evaluating the subexpression(s) is at most $i$. If this is satisfied, typechecking finishes in stage $i+1$. There is a similar constraint on the function application rule (APP), which specifies that the current stage when beginning to evaluate a function be at most the function's starting effect.

\subsubsection{Extensions in Practice}
For clarity, we only present a minimal system here. In practice, the algorithm we implemented for \lang programs differs in two ways beyond simple syntactic differences. First, it performs type and stage \emph{inference}, rather than simply checking the user's annotations, using an imperative algorithm analogous to Algorithm J \cite{milner_1978}.

Our algorithm also allows for polymorphic functions, so that a single function definition can be re-used for different input types or at different starting stages. For example, a function which takes two global variables as arguments and accesses them in order should work for any two arguments where the first is less than the second. To express this, function types can be extended contain constraints on polymorphic stages which appear in the type. These constraints have the form $\epsilon \leq \epsilon$, and can be automatically inferred and checked by the type system.
\clearpage
\section{Soundness of Type System}
In this section we define an operational semantics for the example language defined in Appendix~\ref{app:types}, and prove the soundness of our type system.

\subsection{Operational Semantics}
Our small-step operational semantics is defined on \emph{states} of our program, which are three tuples $(G, n, e)$. Here, $G$ is an array of values such that $G[i]$ is the current value of global variable $g_i$. We write $G[i]$ for the value in $G$ at index $i$, and $G[i := v]$ for the array with all entries the same as $G$, but where index $i$ has value $v$ instead. We say that $G$ is well-typed if $G[i]$ has type $T_i$ for all $i$; that is, if $\emptyset, \epsilon \vdash G[i] : T_i, \epsilon$ for all $\epsilon$.

$n$ is an index into the array indicating the next global variable to be used -- global variables with index less than $n$ are inaccessible. Finally, $e$ is the expression we are evaluating. 

Note the syntactic convention that the metavariable $v$ will only be used to represent expressions which are values. We use a standard definition of variable substitution, where $e[v/x]$ means "e with the value v substituted for the variable x wherever it appears".

\newcommand{\p}{^{\prime}}

\makeunderscoreactive
\begin{figure}[t]
\footnotesize
\noindent\rule{\columnwidth}{.1em}
\begin{mathpar}
	\relationRule{plus-1}{
		(G, n, e_1) \rightarrow (G\p, n\p, e_1\p)
	}{
		(G, n, e_1+e_2) \rightarrow (G\p, n\p, e_1\p + e_2)
	}
	
	\relationRule{plus-2}{
		(G, n, e_2) \rightarrow (G\p, n\p, e_2\p)
	}{
		(G, n, v+e_2) \rightarrow (G\p, n\p, v + e_2\p)
	}
	
	\relationRule{plus-2}{
	    v_1, v_2 \in \mathbb{Z}\\
		\mbox{$v_3$ is the integer sum of $v_1$ and $v_2$}
	}{
		(G, n, v_1+v_2) \rightarrow (G, n, v_3)
	}
	
	\relationRule{let-1}{
		(G, n, e_1) \rightarrow (G\p, n\p, e_1\p)
	}{
		(G, n, \texttt{let } x = e_1 \texttt{ in } e_2) \rightarrow (G\p, n\p, \texttt{let } x = e_1\p \texttt{ in } e_2)
	}
	
	\relationRule{let-2}{
		\ 
	}{
		(G, n, \texttt{let } x = v \texttt{ in } e_2) \rightarrow (G, n,e_2[v/x])
	}
	
	\relationRule{deref-1}{
		(G, n, e) \rightarrow (G\p, n\p, e\p)
	}{
		(G, n, !e) \rightarrow (G\p, n\p, !e\p)
	}

    \relationRule{deref-2}{
        n \leq i
	}{
		(G, n, !g_i) \rightarrow (G, i+1, G[i])
	}

    \relationRule{update-1}{
        (G, n, e_1) \rightarrow (G\p, n\p, e_1\p)
    }{
	    (G, n, e_2 := e_1) \rightarrow (G\p, n\p, e_2 := e_1\p)
    }
    
    \relationRule{update-2}{
        (G, n, e_2) \rightarrow (G\p, n\p, e_2\p)
    }{
	    (G, n, e_2 := v) \rightarrow (G\p, n\p, e_2\p := v)
    }
    
    \relationRule{update-3}{
        n \leq i
    }{
	    (G, n, g_i := v) \rightarrow (G[i := v], i+1, ())
    }
	
	\relationRule{app-1}{
		(G, n, e_1) \rightarrow (G\p, n\p, e_1\p)
	}{
		(G, n, e_1\ e_2) \rightarrow (G\p, n\p, e_1\p\ e_2)
	}
	
	\relationRule{app-2}{
		(G, n, e_2) \rightarrow (G\p, n\p, e_2\p)
	}{
		(G, n, v\ e_2) \rightarrow (G\p, n\p, v\ e_2\p)
	}
	
	\relationRule{app-3}{
		v_1 = \texttt{fun }(x : \tau, \epsilon) \rightarrow e
	}{
		(G, n, v_1\ v_2) \rightarrow (G, n, e[v_2/x])
	}
	
\end{mathpar}
\vspace{-0.5em}
\noindent\rule{\columnwidth}{.1em}%
\vspace{4pt}
\caption{Operational Semantics}
\label{fig:opsem}
\end{figure}
\makeunderscoreletter

\subsection{Important Lemmas}
We first prove a number of useful lemmas. The Canonical Forms lemma says that typing a value does not change the stage, and that only integer values have type integer, and similarly for other types. The Substitution lemma states that uniformly replacing a variable with a value of the same type does not affect our ability to type an expression. Finally, the weakening lemma says that if an expression typechecks starting at some stage, then it also typechecks starting from any earlier stage.

\textbf{Lemma (Canonical forms):} If $\emptyset, \epsilon \vdash v : \tau, \epsilon\p$, then $\epsilon = \epsilon\p$ and:
\begin{itemize}
    \item if $\tau = \texttt{Int}$ then $v \in \Z$
    \item if $\tau = \texttt{ref} (T, k)$, then $v = g_k$ and $T = T_k$.
    \item if $\tau = ({\tau}_{in}, {\epsilon}_{in}) \rightarrow ({\tau}_{out}, {\epsilon}_{out}$ then $v = \texttt{fun } (x : {\tau}_{in}, {\epsilon}_{in}) \rightarrow e$ and $\emptyset[x := {\tau}_{in}], {\epsilon}_{in} \vdash e : {\tau}_{out}, {\epsilon}_{out}$.
\end{itemize}

Proof: Inversion of the typing relation. $\blacksquare$

Definition: If $\Gamma$ is a map, let $\Gamma\backslash x$ denote the same map without a binding for $x$.

\textbf{Lemma (Substitution Lemma):} If $\Gamma[x] = \tau$ and $\emptyset, i \vdash v : \tau, i$ and $\emptyset, \epsilon \vdash e : \tau\p, \epsilon\p$, then $\Gamma\backslash x, \epsilon \vdash e[v/x] : \tau\p, \epsilon\p$

Proof: This is a standard lemma and may be proved for our language in the standard way. $\blacksquare$

\textbf{Lemma (Weakening):} If $\emptyset, \epsilon \vdash e : \tau, \epsilon\p$, and ${\epsilon}_1 \leq \epsilon\p$, then there is some ${\epsilon}_1\p \leq \epsilon\p$ such that $\emptyset, {\epsilon}_1 \vdash e : \tau, {\epsilon}_1\p$.

Proof: Straightforward induction on the typing derivation. $\blacksquare$

\subsection{Progress}
We prove our soundness theorem in the standard way: by combining progress and preservation lemmas.

\textbf{Theorem (Progress):} If $\emptyset, i \vdash e : \tau, j$ then either $e$ is a value or for all well-typed $G$ there exist some $G\p, j\p, e\p$ such that $(G, i, e) \rightarrow (G\p, j\p, e\p)$.

Proof: Structural induction on the typing derivation.

Case INT/UNIT/GLOBAL VARIABLE/ABS: In these cases the expression is already a value, so the result is trivial.

Case LOCAL VARIABLE: This case is impossible, as our typing judgement contains an empty environment.

Case PLUS: In this case, $e = e_1 + e_2$. By induction, either $e_1$ is a value or it steps to some $(G\p, i\p, e_1\p)$. In the latter case, we may apply rule PLUS-1 to show that $(G, i, e_1 + e_2) \rightarrow (G\p, i\p, e_1\p + e_2$.

Similarly, either $e_2$ is a value or it steps to some other $(G\p, i\p, e_2\p)$. If $e_1$ is a value and $e_2$ steps, then we may apply rule PLUS-2 to show that $(G, i, e_1 + e_2) \rightarrow (G\p, i\p, e_1 + e_2\p)$

Finally, if both $e_1$ and $e_2$ are values, then note that by the premises of the PLUS rule, both have type Int. By our canonical forms lemma, $e_1, e_2 \in Z$, so we may apply the PLUS-3 rule to show that $(G, i, e_1 + e_2) \rightarrow (G, i, v)$ where $v$ is the sum of $e_1$ and $e_2$.

Case LET: In this case, $e = \texttt{let } x = e_1 \texttt{ in } e_2$. By induction, either $e_1$ is a value or it steps to something. In the latter case we may apply rule LET-1; otherwise, we may apply rule LET-2.

Case DEREF: In this case, $e = !e_1$. By induction, either $e_1$ is a value or it steps to something. In the latter case we may apply rule DEREF-1; otherwise, note that we have the premise $\emptyset, i \vdash e_1 : \texttt{ref} (T, k), j\p$ where $j\p \leq k$. Since $e_1$ is a value, our canonical forms lemma tells us that $e_1 = G_k$, and $j\p = i$. Hence $i = j\p \leq k$, so we can apply rule DEREF-2 to show that $(G, i, !e_1) \rightarrow (G, k+1, G[k])$.

Case UPDATE: In this case, $e = e_2 := e_1$. As in previous parts, the only interesting case is when $e_1$ and $e_2$ are both values. In that case, as in the DEREF rule, canonical forms tells us that $e_2 = g_k$ for some $k \geq i$, and thus we can apply rule UPDATE-3.

Case APP: In this case, $e = \texttt{let } x = e_1 \texttt{ in } e_2$. The reasoning is analogous to the PLUS case. $\blacksquare$

\subsection{Preservation}

\textbf{Theorem (Preservation):} If $\emptyset, i \vdash e : \tau, j$, $G$ is well-typed, and $(G, i, e) \rightarrow (G\p, i\p, e\p)$, then $G\p$ is also well-typed, and there is some $j\p \leq j$ such that $\emptyset, i\p \vdash e\p : \tau, j\p$.

Proof: Structural induction on the typing derivation.

Case INT/UNIT/GLOBAL VARIABLE: This case cannot occur, because values do not evaluate to anything.

Case LOCAL VARIABLE: This case also cannot occur, because the typing judgement contains an empty environment.

Case PLUS: In this case $e = e_1 + e_2$ and $\tau = \texttt{Int}$. Thus the proof that $e$ steps must have used either PLUS-1, PLUS-2, or PLUS-3. We also have the premises of the PLUS rule: $\emptyset, i \vdash e_1 : \texttt{Int}, k$ and $\emptyset, k \vdash e_2 : \texttt{Int}, j$.
\begin{itemize}
    \item If we used PLUS-1, then we know that $(G, i, e_1) \rightarrow (G\p, i\p, e_1\p)$ and $e\p = e_1\p + e_2$. Thus by induction, $G\p$ is well-typed and $\emptyset, i\p \vdash e_1\p : \texttt{Int}, k\p$ for some $k\p \leq k$. We may use weakening on the second premise to obtain $\emptyset, k\p \vdash e_2 : \texttt{Int}, j\p$ for some $j\p \leq j$, and combine these judgements to show that $\emptyset, i\p \vdash e_1\p + e_2 : \texttt{Int}, j\p$ as required.
    \item If we used PLUS-2, then we know that $(G, i, e_2) \rightarrow (G\p, i\p, e_2\p)$ and $e\p = e_1 + e_2\p$. We also know that $e_1$ is a value, so by canonical forms $e_1 \in \Z$ and $k = i$. Thus we may combine this with the second premise and use induction to conclude that $G\p$ is well-typed, and that $\emptyset, i\p \vdash e_2\p : \texttt{Int}, j\p$ for some $j\p \leq j$. Since $e_1 \in \Z$ we may use the INT rule to show that $\emptyset, i\p \vdash e_1 : \texttt{Int}, j$, and combine this with the previous judgement to show that $\emptyset, i\p \vdash e_1 + e_2\p \vdash \texttt{Int}, j\p$ as required.
    \item If we used PLUS-3, we know that $G\p = G$ and is hence well-typed, $j = i$, and both $e_1$ and $e_2$ are values, so $e_1, e_2 \in \Z$ and thus $e\p \in \Z$ as well. Thus we may simply use the INT rule to show that $\emptyset, i \vdash e\p : \texttt{Int}, j$ as required.
\end{itemize}

Case LET: In this case, $e = \texttt{let } x = e_1 \texttt{ in } e_2$, and we have the premises $\emptyset, i \vdash e_1 : {\tau}_1, k$ and $\emptyset[x := {\tau}_1], k \vdash e_2 : \tau, j$.

The proof that $e$ steps must have used either LET-1 or LET-2. The LET-1 case is analogous to the PLUS-1 case. In the LET-2 case, we know that $e_1$ is a value, $G\p = G$ and hence is well-typed, and $i = j$. By canonical forms, $k = i$. By the substitution lemma, the second premise becomes $\emptyset, i \vdash e_2[e_1/x] : \tau, j$, which is precisely what we wanted to show.

Case DEREF: In this case, $e = !e_1$, and we have the premises that $\emptyset, i \vdash e_1 : \texttt{ref}(T, k), k\p$ where $k\p \leq k$ and $\tau = T$. As usual, we must have used either the DEREF-1 or DEREF-2 rule to prove that $e$ steps. The DEREF-1 case is analogous to the PLUS-1 case.

If we used DEREF-2, then we know that $G\p = G$ is well-typed, $j = k + 1$, $e_1$ is a value, and $e\p = G[k]$. By canonical forms, $e_1 = g_k$ and $\tau = T = T_k$. We need only show that $\emptyset, k+1 \vdash G[k] : \tau, k+1$, which follows immediately from $G$ being well-typed.

Case UPDATE: In this case, $e = e_2 := e_1$, $\tau = \texttt{Unit}$, and we have the premises that $\emptyset, i \vdash e_1 : T, k_1$, $\emptyset, k_1 \vdash e_2 : \texttt{ref}(T, k_2), k_3$ where $k_3 \leq k_2$. As usual, we must have used either UPDATE-1, UPDATE-2, or UPDATE-3 to show that $e$ steps, and the first two cases are analogous to PLUS-1 and PLUS-2, respectively.

In the UPDATE-3 case, we know that the output value is $()$, which can trivially be typed using the UNIT rule. So we need only show that $G\p = G[k_2 := e_1$ is well-typed. But since $e_1$ is a value, we must have used the INT or UNIT rule to prove the first premise, and that rule works for all $\epsilon$. Thus $G\p[k_2]$ has the right type, and all other entries are unchanged, so $G\p$ is well-typed.

Case APP: In this case, $e = e_1\ e_2$, and we have the premises $\emptyset, i \vdash e_1 : ({\tau}_{in}, {\epsilon}_{in}) \rightarrow ({\tau}_{out}, {\epsilon}_{out}), k$ and $\emptyset, k \vdash e_2 : {\tau}_{in}, k_2$ where $k_2 \leq {\epsilon}_{in}$, $\tau = {\tau}_{out}$ and $j = {\epsilon}_{out}$. As usual, we must have used either the APP-1, APP-2, or APP-3 rules here, and the first two cases are again analogous to PLUS-1 and PLUS-2.

If we used the APP-3 rule, then we know that $G\p = G$ is well-typed, that $i\p = i$, that $e_1 = \texttt{fun} (x : {\tau}_1, {\epsilon}_1) \rightarrow e_{body}$ and $e_2$ are both values, and that $e\p = e_{body}[e_2/x]$. Since both $e_1$ and $e_2$ are values, by canonical forms we know that $i = k = k_2$.

By the canonical forms lemma on $e_1$, we know that $\emptyset[x := {\tau}_{in}], {\epsilon}_{in} \vdash e_{body} : {\tau}_{out}, {\epsilon}_{out}$. Now, $e_2$ is a value, and by the second premise it has type ${\tau}_{in}$; thus by the substitution lemma $\emptyset, {\epsilon}_{in} \vdash e_{body}[e_2/x] : {\tau}_{out}, {\epsilon}_{out}$. Since $i = k_2 \leq {\epsilon}_{in}$, by weakening there is some ${\epsilon}_{out}\p$ such that $\emptyset, i \vdash e_{body}[e_2/x] : {\tau}_{out}, {\epsilon}_{out}\p$ $\blacksquare$

\subsection{Soundness}
Finally, we combine the progress and preservation theorems to prove that expressions which typecheck always evaluate -- that is, "Well-typed programs do not get stuck". We denote the transitive closure of the evaluation relation by $\rightarrow^*$.

\textbf{Theorem (Soundness):} If $\emptyset, {\epsilon}_1 \vdash e : \tau, {\epsilon}_2$, then either $e$ is a value, or $e \rightarrow e^{\prime}$ and there is some ${\epsilon}_1^{\prime}$ such that $\emptyset, {\epsilon}_1^{\prime} \vdash e^{\prime} : \tau, {\epsilon}_2$.

Proof: Inductively apply preservation to show that $G_2$ and $e_2$ are well-typed, then apply progress to show that $e_2$ is either a value, or another step can be taken. $\blacksquare$

%% file: sections/appendixmemop.tex
\section{Memop Limitations}
\label{app:memop}

\begin{figure}[h]
\begin{lstlisting}[style=DptBlock]
memop compoundCondition(int memval, int y){
  if (memval == 1 || memval == 2) {
    return memval;
  } else {
    return y;
  }
}
memop twoLocalArgs(int memval, int y, int z){
  if (memval == 1) {
    return y;
  } else {
    return z;
  }
}
const int N = 10;
memop multipy(int memval, int x){
    return (N * memval) + x;
}
\end{lstlisting}
\caption{Memops that are invalid because of: 1) compound conditional expressions; 2) accessing too much local state (\IE, packet header or metadata) and; 3) arithmetic operations that are too complex.}
\label{fig:badmemops}
\end{figure}

\rerevised{
This appendix discusses stateful operations that can be implemented by the Tofino, but are not supported by \lang's base \emph{memop} syntax. Ultimately, \emph{memops} rule out some implementable operations to provide a uniform base abstraction where: 1) every array method call with valid \emph{memops} is guaranteed to be implementable by a stateful ALU; 2) any \emph{memop} can be used in any array method; and 3) all \emph{memops} have the same syntactic restrictions.

A uniform \emph{memop} abstraction reduces completeness because some array methods place more restrictions on \emph{memops} than others. Specifically, while each array method call compiles to a single stateful ALU instruction, an \inl{Array.update} call (\IE, a parallel get and set) requires us to compile two \emph{memops} to a single instruction, whereas \inl{Array.get} and \inl{Array.set} only require us to compile one \emph{memop} to the instruction. Thus, compared to an \inl{Array.get} or \inl{Array.set}, each memop that we pass to \inl{Array.update} can only safely use ``half'' of the stateful ALU's capabilities. Since we want any \emph{memop} to be usable in any array method, the base \emph{memop} syntax must reflect this constraint. This, in turn, disallows some of the more complicated set or get operations that the Tofino could implement.

Figure~\ref{fig:badmemops} gives three examples of \emph{memops} that are not valid in Lucid, but can be implemented on the Tofino. Each example \emph{memop} could be implemented in a stateful ALU, but would not leave enough stateful ALU resources to guarantee that a second \emph{memop} of \inl{array.update} could also fit. Each example stresses a different kind of primitive within the stateful ALU. All of these examples could be supported by a future version of Lucid with a special kind of \inl{memop} that is not allowed to be used by the fully general version of \inl{Array.update}.}

%% file: sections/appendixartifact.tex

\section{Artifact Appendix}

\subsection{Abstract}

The artifact associated with this paper is the \lang compiler, which translates 
\lang programs into Tofino-optimized \texttt{P4\_16}. As of publication, we are actively developing the \lang compiler and using \lang to build data-plane 
applications for several other projects. 

\subsection{Scope}

The \lang compiler is intended for general Tofino programming. We believe it can 
significantly reduce programmer effort for a wide range of data-plane applications. The \lang interpreter, which can also be found in this repository, enables rapid prototyping and testing of data-plane applications without requiring access to the Tofino toolchain. 

Additionally, the artifact can be used to reproduce Figure~\ref{tab:usecaseeval} from the evaluation. 

\subsection{Contents}

The main branch of this repository contains the \lang compiler, the \lang interpreter, example applications, usage instructions and tutorials, and scripts for automating deployment to P4. 

The \texttt{sigcomm21_artifact} branch of the \lang repository contains a snapshot of the \lang compiler from 5/2021 with instructions to reproduce Figure~\ref{tab:usecaseeval} from the evaluation. 

\subsection{Hosting}

The repository is hosted on GitHub, at:\\\url{https://github.com/PrincetonUniversity/lucid}. 

\noindent The branch to reproduce Figure~\ref{tab:usecaseeval} is at:\\ 
\begin{small}
\url{https://github.com/PrincetonUniversity/lucid/tree/sigcomm21_artifact}
\end{small}.

\subsection{Requirements}

The repository branch associated with Figure~\ref{tab:usecaseeval} was tested with: virtualbox 6.1.8 (\url{https://www.virtualbox.org/wiki/Downloads}); Vagrant 2.2.9 (\url{https://www.vagrantup.com/downloads}); and the Intel P4 studio SDE version 9.5.0. P4 studio is only necessary if you wish to compile the output of the \lang compiler to the Tofino. 

The main branch of the \lang repository lists current requirements in its \texttt{readme.md}.


%% file: paper.bbl

\begin{thebibliography}{35}


\ifx \showCODEN    \undefined \def \showCODEN     #1{\unskip}     \fi
\ifx \showDOI      \undefined \def \showDOI       #1{#1}\fi
\ifx \showISBNx    \undefined \def \showISBNx     #1{\unskip}     \fi
\ifx \showISBNxiii \undefined \def \showISBNxiii  #1{\unskip}     \fi
\ifx \showISSN     \undefined \def \showISSN      #1{\unskip}     \fi
\ifx \showLCCN     \undefined \def \showLCCN      #1{\unskip}     \fi
\ifx \shownote     \undefined \def \shownote      #1{#1}          \fi
\ifx \showarticletitle \undefined \def \showarticletitle #1{#1}   \fi
\ifx \showURL      \undefined \def \showURL       {\relax}        \fi
\providecommand\bibfield[2]{#2}
\providecommand\bibinfo[2]{#2}
\providecommand\natexlab[1]{#1}
\providecommand\showeprint[2][]{arXiv:#2}

\bibitem[\protect\citeauthoryear{Alizadeh, Edsall, Dharmapurikar, Vaidyanathan,
  Chu, Fingerhut, Lam, Matus, Pan, Yadav, and Varghese}{Alizadeh
  et~al\mbox{.}}{2014}]%
        {alizadeh2014conga}
\bibfield{author}{\bibinfo{person}{Mohammad Alizadeh}, \bibinfo{person}{Tom
  Edsall}, \bibinfo{person}{Sarang Dharmapurikar}, \bibinfo{person}{Ramanan
  Vaidyanathan}, \bibinfo{person}{Kevin Chu}, \bibinfo{person}{Andy Fingerhut},
  \bibinfo{person}{The~Vinh Lam}, \bibinfo{person}{Francis Matus},
  \bibinfo{person}{Rong Pan}, \bibinfo{person}{Navindra Yadav}, {and}
  \bibinfo{person}{George Varghese}.} \bibinfo{year}{2014}\natexlab{}.
\newblock \showarticletitle{CONGA: Distributed congestion-aware load balancing
  for datacenters}. In \bibinfo{booktitle}{\emph{ACM SIGCOMM}}.
  \bibinfo{pages}{503--514}.
\newblock


\bibitem[\protect\citeauthoryear{Anderson, Foster, Guha, Jeannin, Kozen,
  Schlesinger, and Walker}{Anderson et~al\mbox{.}}{2014}]%
        {netkat}
\bibfield{author}{\bibinfo{person}{Carolyn~Jane Anderson},
  \bibinfo{person}{Nate Foster}, \bibinfo{person}{Arjun Guha},
  \bibinfo{person}{Jean-Baptiste Jeannin}, \bibinfo{person}{Dexter Kozen},
  \bibinfo{person}{Cole Schlesinger}, {and} \bibinfo{person}{David Walker}.}
  \bibinfo{year}{2014}\natexlab{}.
\newblock \showarticletitle{{NetKAT}: Semantic Foundations for Networks}. In
  \bibinfo{booktitle}{\emph{ACM SIGPLAN-SIGACT Symposium on Principles of
  Programming Languages}}. \bibinfo{pages}{113–126}.
\newblock


\bibitem[\protect\citeauthoryear{Arbitman, Naor, and Segev}{Arbitman
  et~al\mbox{.}}{2009}]%
        {cuckoostashprovable}
\bibfield{author}{\bibinfo{person}{Yuriy Arbitman}, \bibinfo{person}{Moni
  Naor}, {and} \bibinfo{person}{Gil Segev}.} \bibinfo{year}{2009}\natexlab{}.
\newblock \showarticletitle{De-amortized cuckoo hashing: Provable worst-case
  performance and experimental results}. In
  \bibinfo{booktitle}{\emph{International Colloquium on Automata, Languages,
  and Programming}}. Springer, \bibinfo{pages}{107--118}.
\newblock


\bibitem[\protect\citeauthoryear{Berde, Gerola, Hart, Higuchi, Kobayashi,
  Koide, Lantz, O'Connor, Radoslavov, Snow, et~al\mbox{.}}{Berde
  et~al\mbox{.}}{2014}]%
        {onos}
\bibfield{author}{\bibinfo{person}{Pankaj Berde}, \bibinfo{person}{Matteo
  Gerola}, \bibinfo{person}{Jonathan Hart}, \bibinfo{person}{Yuta Higuchi},
  \bibinfo{person}{Masayoshi Kobayashi}, \bibinfo{person}{Toshio Koide},
  \bibinfo{person}{Bob Lantz}, \bibinfo{person}{Brian O'Connor},
  \bibinfo{person}{Pavlin Radoslavov}, \bibinfo{person}{William Snow},
  {et~al\mbox{.}}} \bibinfo{year}{2014}\natexlab{}.
\newblock \showarticletitle{ONOS: Towards an open, distributed SDN OS}. In
  \bibinfo{booktitle}{\emph{Workshop on Hot Topics in Software Defined
  Networking}}. \bibinfo{pages}{1--6}.
\newblock


\bibitem[\protect\citeauthoryear{Bosshart, Gibb, Kim, Varghese, McKeown,
  Izzard, Mujica, and Horowitz}{Bosshart et~al\mbox{.}}{2013}]%
        {bosshart2013forwarding}
\bibfield{author}{\bibinfo{person}{Pat Bosshart}, \bibinfo{person}{Glen Gibb},
  \bibinfo{person}{Hun-Seok Kim}, \bibinfo{person}{George Varghese},
  \bibinfo{person}{Nick McKeown}, \bibinfo{person}{Martin Izzard},
  \bibinfo{person}{Fernando Mujica}, {and} \bibinfo{person}{Mark Horowitz}.}
  \bibinfo{year}{2013}\natexlab{}.
\newblock \showarticletitle{Forwarding metamorphosis: Fast programmable
  match-action processing in hardware for SDN}. In
  \bibinfo{booktitle}{\emph{ACM SIGCOMM}}. \bibinfo{pages}{99--110}.
\newblock


\bibitem[\protect\citeauthoryear{Chen}{Chen}{2020}]%
        {chen2020implementing}
\bibfield{author}{\bibinfo{person}{Xiaoqi Chen}.}
  \bibinfo{year}{2020}\natexlab{}.
\newblock \showarticletitle{Implementing AES encryption on programmable
  switches via scrambled lookup tables}. In \bibinfo{booktitle}{\emph{ACM
  SIGCOMM Workshop on Secure Programmable Network Infrastructure}}.
  \bibinfo{pages}{8--14}.
\newblock


\bibitem[\protect\citeauthoryear{DeLine and Fahndrich}{DeLine and
  Fahndrich}{1999}]%
        {vault}
\bibfield{author}{\bibinfo{person}{Rob DeLine} {and} \bibinfo{person}{Manuel
  Fahndrich}.} \bibinfo{year}{1999}\natexlab{}.
\newblock \showarticletitle{Natural deduction for intuitionistic
  non-commutative linear logic}. In \bibinfo{booktitle}{\emph{International
  Conference on Typed Lambda Calculi and Applications}}.
\newblock


\bibitem[\protect\citeauthoryear{Foster, Harrison, Freedman, Monsanto, Rexford,
  Story, and Walker}{Foster et~al\mbox{.}}{2011}]%
        {frenetic}
\bibfield{author}{\bibinfo{person}{Nate Foster}, \bibinfo{person}{Rob
  Harrison}, \bibinfo{person}{Michael~J. Freedman},
  \bibinfo{person}{Christopher Monsanto}, \bibinfo{person}{Jennifer Rexford},
  \bibinfo{person}{Alec Story}, {and} \bibinfo{person}{David Walker}.}
  \bibinfo{year}{2011}\natexlab{}.
\newblock \showarticletitle{Frenetic: A Network Programming Language}. In
  \bibinfo{booktitle}{\emph{ACM International Conference on Functional
  Programming}}. \bibinfo{pages}{279–291}.
\newblock


\bibitem[\protect\citeauthoryear{Gao, Zhai, Liu, Miao, Zhou, Tian, Sun, Cai,
  Zhang, and Yu}{Gao et~al\mbox{.}}{2020b}]%
        {lyra}
\bibfield{author}{\bibinfo{person}{Jiaqi Gao}, \bibinfo{person}{Ennan Zhai},
  \bibinfo{person}{Hongqiang~Harry Liu}, \bibinfo{person}{Rui Miao},
  \bibinfo{person}{Yu Zhou}, \bibinfo{person}{Bingchuan Tian},
  \bibinfo{person}{Chen Sun}, \bibinfo{person}{Dennis Cai},
  \bibinfo{person}{Ming Zhang}, {and} \bibinfo{person}{Minlan Yu}.}
  \bibinfo{year}{2020}\natexlab{b}.
\newblock \showarticletitle{Lyra: A Cross-Platform Language and Compiler for
  Data Plane Programming on Heterogeneous ASICs}. In
  \bibinfo{booktitle}{\emph{ACM SIGCOMM}}. \bibinfo{pages}{435–450}.
\newblock


\bibitem[\protect\citeauthoryear{Gao, Kim, Wong, Raghunathan, Varma, Kannan,
  Sivaraman, Narayana, and Gupta}{Gao et~al\mbox{.}}{2020a}]%
        {chipmunk}
\bibfield{author}{\bibinfo{person}{Xiangyu Gao}, \bibinfo{person}{Taegyun Kim},
  \bibinfo{person}{Michael~D. Wong}, \bibinfo{person}{Divya Raghunathan},
  \bibinfo{person}{Aatish~Kishan Varma}, \bibinfo{person}{Pravein~Govindan
  Kannan}, \bibinfo{person}{Anirudh Sivaraman}, \bibinfo{person}{Srinivas
  Narayana}, {and} \bibinfo{person}{Aarti Gupta}.}
  \bibinfo{year}{2020}\natexlab{a}.
\newblock \showarticletitle{Switch Code Generation Using Program Synthesis}. In
  \bibinfo{booktitle}{\emph{ACM SIGCOMM}}. \bibinfo{pages}{44–61}.
\newblock


\bibitem[\protect\citeauthoryear{Group}{Group}{[n.d.]}]%
        {p4runtime}
\bibfield{author}{\bibinfo{person}{The P4.org API~Working Group}.}
  \bibinfo{year}{[n.d.]}\natexlab{}.
\newblock \bibinfo{title}{{P4Runtime} {Specification}}.
\newblock
\newblock
\urldef\tempurl%
\url{https://p4lang.github.io/p4runtime/spec/main/P4Runtime-Spec.html}
\showURL{%
\tempurl}


\bibitem[\protect\citeauthoryear{Gupta and McKeown}{Gupta and McKeown}{2001}]%
        {gupta2001algorithms}
\bibfield{author}{\bibinfo{person}{Pankaj Gupta} {and} \bibinfo{person}{Nick
  McKeown}.} \bibinfo{year}{2001}\natexlab{}.
\newblock \showarticletitle{Algorithms for packet classification}.
\newblock \bibinfo{journal}{\emph{IEEE Network}} \bibinfo{volume}{15},
  \bibinfo{number}{2} (\bibinfo{year}{2001}), \bibinfo{pages}{24--32}.
\newblock


\bibitem[\protect\citeauthoryear{Heller, Seetharaman, Mahadevan, Yiakoumis,
  Sharma, Banerjee, and McKeown}{Heller et~al\mbox{.}}{2010}]%
        {elastictree}
\bibfield{author}{\bibinfo{person}{Brandon Heller}, \bibinfo{person}{Srinivasan
  Seetharaman}, \bibinfo{person}{Priya Mahadevan}, \bibinfo{person}{Yiannis
  Yiakoumis}, \bibinfo{person}{Puneet Sharma}, \bibinfo{person}{Sujata
  Banerjee}, {and} \bibinfo{person}{Nick McKeown}.}
  \bibinfo{year}{2010}\natexlab{}.
\newblock \showarticletitle{{ElasticTree}: Saving energy in data center
  networks}. In \bibinfo{booktitle}{\emph{USENIX Networked Systems Design and
  Implementation}}, Vol.~\bibinfo{volume}{10}. \bibinfo{pages}{249--264}.
\newblock


\bibitem[\protect\citeauthoryear{Hogan, Landau-Feibish, Tahmasbi~Arashloo,
  Rexford, Walker, and Harrison}{Hogan et~al\mbox{.}}{2020}]%
        {p4all}
\bibfield{author}{\bibinfo{person}{Mary Hogan}, \bibinfo{person}{Shir
  Landau-Feibish}, \bibinfo{person}{Mina Tahmasbi~Arashloo},
  \bibinfo{person}{Jennifer Rexford}, \bibinfo{person}{David Walker}, {and}
  \bibinfo{person}{Rob Harrison}.} \bibinfo{year}{2020}\natexlab{}.
\newblock \showarticletitle{Elastic Switch Programming with P4All}. In
  \bibinfo{booktitle}{\emph{ACM SIGCOMM HotNets Networks}}.
  \bibinfo{pages}{168–174}.
\newblock


\bibitem[\protect\citeauthoryear{Hsu, Beckett, Chen, Rexford, and Walker}{Hsu
  et~al\mbox{.}}{2020}]%
        {hsu2020contra}
\bibfield{author}{\bibinfo{person}{Kuo-Feng Hsu}, \bibinfo{person}{Ryan
  Beckett}, \bibinfo{person}{Ang Chen}, \bibinfo{person}{Jennifer Rexford},
  {and} \bibinfo{person}{David Walker}.} \bibinfo{year}{2020}\natexlab{}.
\newblock \showarticletitle{Contra: A programmable system for performance-aware
  routing}. In \bibinfo{booktitle}{\emph{USENIX Symposium on Networked Systems
  Design and Implementation}}. \bibinfo{pages}{701--721}.
\newblock


\bibitem[\protect\citeauthoryear{Ibanez, Antichi, Brebner, and McKeown}{Ibanez
  et~al\mbox{.}}{2019}]%
        {ibanez2019event}
\bibfield{author}{\bibinfo{person}{Stephen Ibanez}, \bibinfo{person}{Gianni
  Antichi}, \bibinfo{person}{Gordon Brebner}, {and} \bibinfo{person}{Nick
  McKeown}.} \bibinfo{year}{2019}\natexlab{}.
\newblock \showarticletitle{Event-driven packet processing}. In
  \bibinfo{booktitle}{\emph{ACM Workshop on Hot Topics in Networks}}.
  \bibinfo{pages}{133--140}.
\newblock


\bibitem[\protect\citeauthoryear{Igarashi and Kobayashi}{Igarashi and
  Kobayashi}{2001}]%
        {igarashi+:resource-usage-analysis}
\bibfield{author}{\bibinfo{person}{Atsushi Igarashi} {and}
  \bibinfo{person}{Naoki Kobayashi}.} \bibinfo{year}{2001}\natexlab{}.
\newblock \showarticletitle{Enforcing high-level protocols in low-level
  software}.
\newblock \bibinfo{journal}{\emph{SIGPLAN Notices}}  \bibinfo{volume}{36}
  (\bibinfo{date}{May} \bibinfo{year}{2001}).
\newblock
Issue 5.


\bibitem[\protect\citeauthoryear{Katta, Hira, Kim, Sivaraman, and
  Rexford}{Katta et~al\mbox{.}}{2016}]%
        {katta2016hula}
\bibfield{author}{\bibinfo{person}{Naga Katta}, \bibinfo{person}{Mukesh Hira},
  \bibinfo{person}{Changhoon Kim}, \bibinfo{person}{Anirudh Sivaraman}, {and}
  \bibinfo{person}{Jennifer Rexford}.} \bibinfo{year}{2016}\natexlab{}.
\newblock \showarticletitle{Hula: Scalable load balancing using programmable
  data planes}. In \bibinfo{booktitle}{\emph{ACM SIGCOMM Symposium on SDN
  Research}}. \bibinfo{pages}{1--12}.
\newblock


\bibitem[\protect\citeauthoryear{Kirsch, Mitzenmacher, and Wieder}{Kirsch
  et~al\mbox{.}}{2008}]%
        {cuckoostash}
\bibfield{author}{\bibinfo{person}{Adam Kirsch}, \bibinfo{person}{Michael
  Mitzenmacher}, {and} \bibinfo{person}{Udi Wieder}.}
  \bibinfo{year}{2008}\natexlab{}.
\newblock \showarticletitle{More robust hashing: Cuckoo hashing with a stash}.
  In \bibinfo{booktitle}{\emph{European Symposium on Algorithms}}. Springer,
  \bibinfo{pages}{611--622}.
\newblock


\bibitem[\protect\citeauthoryear{Liu, Halperin, Krishnamurthy, and
  Anderson}{Liu et~al\mbox{.}}{2013}]%
        {f10}
\bibfield{author}{\bibinfo{person}{Vincent Liu}, \bibinfo{person}{Daniel
  Halperin}, \bibinfo{person}{Arvind Krishnamurthy}, {and}
  \bibinfo{person}{Thomas Anderson}.} \bibinfo{year}{2013}\natexlab{}.
\newblock \showarticletitle{F10: A Fault-Tolerant Engineered Network}. In
  \bibinfo{booktitle}{\emph{USENIX Symposium on Networked Systems Design and
  Implementation}}. \bibinfo{pages}{399--412}.
\newblock


\bibitem[\protect\citeauthoryear{Liu, Namkung, Nikolaidis, Lee, Kim, Jin,
  Braverman, Yu, and Sekar}{Liu et~al\mbox{.}}{2021}]%
        {jaqen}
\bibfield{author}{\bibinfo{person}{Zaoxing Liu}, \bibinfo{person}{Hun Namkung},
  \bibinfo{person}{Georgios Nikolaidis}, \bibinfo{person}{Jeongkeun Lee},
  \bibinfo{person}{Changhoon Kim}, \bibinfo{person}{Xin Jin},
  \bibinfo{person}{Vladimir Braverman}, \bibinfo{person}{Minlan Yu}, {and}
  \bibinfo{person}{Vyas Sekar}.} \bibinfo{year}{2021}\natexlab{}.
\newblock \showarticletitle{Jaqen: A High-Performance Switch-Native Approach
  for Detecting and Mitigating Volumetric DDoS Attacks with Programmable
  Switches}. In \bibinfo{booktitle}{\emph{USENIX Security Symposium}}.
\newblock


\bibitem[\protect\citeauthoryear{Milner}{Milner}{1978}]%
        {milner_1978}
\bibfield{author}{\bibinfo{person}{Robin Milner}.}
  \bibinfo{year}{1978}\natexlab{}.
\newblock \showarticletitle{A theory of type polymorphism in programming}.
\newblock \bibinfo{journal}{\emph{J. Comput. System Sci.}}
  \bibinfo{volume}{17}, \bibinfo{number}{3} (\bibinfo{year}{1978}),
  \bibinfo{pages}{348–375}.
\newblock
\urldef\tempurl%
\url{https://doi.org/10.1016/0022-0000(78)90014-4}
\showDOI{\tempurl}


\bibitem[\protect\citeauthoryear{Nelson, Ferguson, Scheer, and
  Krishnamurthi}{Nelson et~al\mbox{.}}{2014}]%
        {flowlog}
\bibfield{author}{\bibinfo{person}{Tim Nelson}, \bibinfo{person}{Andrew~D.
  Ferguson}, \bibinfo{person}{Michael~J.G. Scheer}, {and}
  \bibinfo{person}{Shriram Krishnamurthi}.} \bibinfo{year}{2014}\natexlab{}.
\newblock \showarticletitle{Tierless Programming and Reasoning for
  Software-Defined Networks}. In \bibinfo{booktitle}{\emph{USENIX Networked
  Systems Design and Implementation}}. \bibinfo{pages}{519--531}.
\newblock


\bibitem[\protect\citeauthoryear{Neugebauer, Antichi, Zazo, Audzevich,
  L{\'o}pez-Buedo, and Moore}{Neugebauer et~al\mbox{.}}{2018}]%
        {pcieperf}
\bibfield{author}{\bibinfo{person}{Rolf Neugebauer}, \bibinfo{person}{Gianni
  Antichi}, \bibinfo{person}{Jos{\'e}~Fernando Zazo}, \bibinfo{person}{Yury
  Audzevich}, \bibinfo{person}{Sergio L{\'o}pez-Buedo}, {and}
  \bibinfo{person}{Andrew~W Moore}.} \bibinfo{year}{2018}\natexlab{}.
\newblock \showarticletitle{Understanding PCIe performance for end host
  networking}. In \bibinfo{booktitle}{\emph{ACM SIGCOMM}}. ACM,
  \bibinfo{pages}{327--341}.
\newblock


\bibitem[\protect\citeauthoryear{Pagh and Rodler}{Pagh and Rodler}{2004}]%
        {pagh2004cuckoo}
\bibfield{author}{\bibinfo{person}{Rasmus Pagh} {and}
  \bibinfo{person}{Flemming~Friche Rodler}.} \bibinfo{year}{2004}\natexlab{}.
\newblock \showarticletitle{Cuckoo hashing}.
\newblock \bibinfo{journal}{\emph{Journal of Algorithms}} \bibinfo{volume}{51},
  \bibinfo{number}{2} (\bibinfo{year}{2004}), \bibinfo{pages}{122--144}.
\newblock


\bibitem[\protect\citeauthoryear{Polakow and Pfenning}{Polakow and
  Pfenning}{1999}]%
        {polakow-ordered}
\bibfield{author}{\bibinfo{person}{Jeff Polakow} {and} \bibinfo{person}{Frank
  Pfenning}.} \bibinfo{year}{1999}\natexlab{}.
\newblock \showarticletitle{Natural deduction for intuitionistic
  non-commutative linear logic}. In \bibinfo{booktitle}{\emph{International
  Conference on Typed Lambda Calculi and Applications}}.
\newblock


\bibitem[\protect\citeauthoryear{Shah, Kumar, Vutukuru, and Kulkarni}{Shah
  et~al\mbox{.}}{2020}]%
        {turboepc}
\bibfield{author}{\bibinfo{person}{Rinku Shah}, \bibinfo{person}{Vikas Kumar},
  \bibinfo{person}{Mythili Vutukuru}, {and} \bibinfo{person}{Purushottam
  Kulkarni}.} \bibinfo{year}{2020}\natexlab{}.
\newblock \showarticletitle{TurboEPC: Leveraging Dataplane Programmability to
  Accelerate the Mobile Packet Core}. In \bibinfo{booktitle}{\emph{ACM
  Symposium on SDN Research}}. \bibinfo{pages}{83–95}.
\newblock


\bibitem[\protect\citeauthoryear{Sivaraman, Cheung, Budiu, Kim, Alizadeh,
  Balakrishnan, Varghese, McKeown, and Licking}{Sivaraman
  et~al\mbox{.}}{2016}]%
        {domino}
\bibfield{author}{\bibinfo{person}{Anirudh Sivaraman}, \bibinfo{person}{Alvin
  Cheung}, \bibinfo{person}{Mihai Budiu}, \bibinfo{person}{Changhoon Kim},
  \bibinfo{person}{Mohammad Alizadeh}, \bibinfo{person}{Hari Balakrishnan},
  \bibinfo{person}{George Varghese}, \bibinfo{person}{Nick McKeown}, {and}
  \bibinfo{person}{Steve Licking}.} \bibinfo{year}{2016}\natexlab{}.
\newblock \showarticletitle{Packet transactions: High-level programming for
  line-rate switches}. In \bibinfo{booktitle}{\emph{ACM SIGCOMM}}.
  \bibinfo{pages}{15--28}.
\newblock


\bibitem[\protect\citeauthoryear{Smolka, Kumar, Kahn, Foster, Hsu, Kozen, and
  Silva}{Smolka et~al\mbox{.}}{2019}]%
        {mcnetkat}
\bibfield{author}{\bibinfo{person}{Steffen Smolka}, \bibinfo{person}{Praveen
  Kumar}, \bibinfo{person}{David~M. Kahn}, \bibinfo{person}{Nate Foster},
  \bibinfo{person}{Justin Hsu}, \bibinfo{person}{Dexter Kozen}, {and}
  \bibinfo{person}{Alexandra Silva}.} \bibinfo{year}{2019}\natexlab{}.
\newblock \showarticletitle{Scalable Verification of Probabilistic Networks}.
  In \bibinfo{booktitle}{\emph{ACM SIGPLAN Programming Language Design and
  Implementation}}. \bibinfo{pages}{190–203}.
\newblock


\bibitem[\protect\citeauthoryear{Sonchack, Michel, Aviv, Keller, and
  Smith}{Sonchack et~al\mbox{.}}{2018}]%
        {starflow}
\bibfield{author}{\bibinfo{person}{John Sonchack}, \bibinfo{person}{Oliver
  Michel}, \bibinfo{person}{Adam~J Aviv}, \bibinfo{person}{Eric Keller}, {and}
  \bibinfo{person}{Jonathan~M Smith}.} \bibinfo{year}{2018}\natexlab{}.
\newblock \showarticletitle{Scaling hardware accelerated network monitoring to
  concurrent and dynamic queries with {*Flow}}. In
  \bibinfo{booktitle}{\emph{USENIX Annual Technical Conference}}.
  \bibinfo{pages}{823--835}.
\newblock


\bibitem[\protect\citeauthoryear{Swamy, Rucker, Shahbaz, and Olukotun}{Swamy
  et~al\mbox{.}}{2020}]%
        {swamy2020taurus}
\bibfield{author}{\bibinfo{person}{Tushar Swamy}, \bibinfo{person}{Alexander
  Rucker}, \bibinfo{person}{Muhammad Shahbaz}, {and} \bibinfo{person}{Kunle
  Olukotun}.} \bibinfo{year}{2020}\natexlab{}.
\newblock \showarticletitle{Taurus: An intelligent data plane}.
\newblock \bibinfo{journal}{\emph{arXiv preprint arXiv:2002.08987}}
  (\bibinfo{year}{2020}).
\newblock


\bibitem[\protect\citeauthoryear{Vass, B{\'e}rczi-Kov{\'a}cs, Raiciu, and
  R{\'e}tv{\'a}ri}{Vass et~al\mbox{.}}{2020}]%
        {vass2020compiling}
\bibfield{author}{\bibinfo{person}{Bal{\'a}zs Vass}, \bibinfo{person}{Erika
  B{\'e}rczi-Kov{\'a}cs}, \bibinfo{person}{Costin Raiciu}, {and}
  \bibinfo{person}{G{\'a}bor R{\'e}tv{\'a}ri}.}
  \bibinfo{year}{2020}\natexlab{}.
\newblock \showarticletitle{Compiling Packet Programs to Reconfigurable
  Switches: Theory and Algorithms}. In \bibinfo{booktitle}{\emph{P4 Workshop in
  Europe}}. \bibinfo{pages}{28--35}.
\newblock


\bibitem[\protect\citeauthoryear{Yasukata, Honda, Santry, and Eggert}{Yasukata
  et~al\mbox{.}}{2016}]%
        {stackmap}
\bibfield{author}{\bibinfo{person}{Kenichi Yasukata}, \bibinfo{person}{Michio
  Honda}, \bibinfo{person}{Douglas Santry}, {and} \bibinfo{person}{Lars
  Eggert}.} \bibinfo{year}{2016}\natexlab{}.
\newblock \showarticletitle{StackMap: Low-Latency Networking with the {OS}
  Stack and Dedicated NICs}. In \bibinfo{booktitle}{\emph{USENIX Annual
  Technical Conference}}. \bibinfo{pages}{43--56}.
\newblock


\bibitem[\protect\citeauthoryear{Yu, Sonchack, and Liu}{Yu
  et~al\mbox{.}}{2020}]%
        {mantis}
\bibfield{author}{\bibinfo{person}{Liangcheng Yu}, \bibinfo{person}{John
  Sonchack}, {and} \bibinfo{person}{Vincent Liu}.}
  \bibinfo{year}{2020}\natexlab{}.
\newblock \showarticletitle{Mantis: Reactive Programmable Switches}. In
  \bibinfo{booktitle}{\emph{ACM SIGCOMM}}. \bibinfo{pages}{296--309}.
\newblock


\bibitem[\protect\citeauthoryear{Zeno, Ports, Nelson, and Silberstein}{Zeno
  et~al\mbox{.}}{2020}]%
        {10.1145/3422604.3425946}
\bibfield{author}{\bibinfo{person}{Lior Zeno}, \bibinfo{person}{Dan R.~K.
  Ports}, \bibinfo{person}{Jacob Nelson}, {and} \bibinfo{person}{Mark
  Silberstein}.} \bibinfo{year}{2020}\natexlab{}.
\newblock \showarticletitle{{SwiShmem}: Distributed Shared State Abstractions
  for Programmable Switches}. In \bibinfo{booktitle}{\emph{ACM SIGCOMM HotNets
  Workshop}}.
\newblock


\end{thebibliography}
